\documentclass[fleqn,usenatbib]{mnras}
\usepackage[T1]{fontenc}

\DeclareRobustCommand{\VAN}[3]{#2}
\let\VANthebibliography\thebibliography
\def\thebibliography{\DeclareRobustCommand{\VAN}[3]{##3}\VANthebibliography}

\usepackage{graphicx}
\usepackage{amsmath}
\usepackage{amssymb}

\usepackage{rotating}
\usepackage{xcolor}
\usepackage{bm}
\usepackage{soul}

\definecolor{notecolor}{rgb}{0.8,0,0}
\definecolor{color2}{rgb}{0.0,0.8,0.0}

\title[Cosmological initial conditions]{Particle initialization effects on Lyman-$\bm{\alpha}$ forest statistics in cosmological SPH simulations}

\newcommand{\lya}{Lyman-$\alpha$}
\newcommand{\ud}{\mathrm{d}}
\newcommand{\ngenic}{NGenIC}
\newcommand{\tlpt}{\textsc{2LPTic}}
\newcommand{\cosmic}{CosmicIC}
\newcommand{\mpgenic}{MP-GenIC}
\newcommand{\music}{MUSIC}
\newcommand{\gadget}{P-Gadget3}

\author[Khan et al.]{{Nabendu Kumar Khan,$^{1}$\thanks{E-mail: nabendu.khan@tifr.res.in}
  Girish Kulkarni,$^{1}$
  James S.~Bolton,$^{2}$
  Martin G.~Haehnelt,$^{3,4}$}
\newauthor{Vid Ir\v{s}i\v{c},$^{4}$ Ewald Puchwein$^5$, and Shikhar Asthana$^{3,4}$}\\
  $^{1}$Tata Institute of Fundamental Research, Homi Bhabha Road, Mumbai 400005, India\\
  $^{2}$School of Physics and Astronomy, University of Nottingham, University Park, Nottingham NG7 2RD, UK\\
  $^3$Institute of Astronomy, University of Cambridge, Madingley Road, Cambridge CB3 0HA, UK \\
  $^4$Kavli Institute of Cosmology, University of Cambridge, Madingley Road, Cambridge CB3 0HA, UK\\
  $^5$Leibniz-Institut f\"ur Astrophysik Potsdam, An der Sternwarte 16, D14482 Potsdam, Germany 
}

\date{Accepted ---. Received ---; in original form ---}

\pubyear{2022}

\begin{document}
\label{firstpage}
\pagerange{\pageref{firstpage}--\pageref{lastpage}}
\maketitle

\begin{abstract}
Confronting measurements of the \lya\ forest with cosmological hydrodynamical simulations has produced stringent constraints on models of particle dark matter and the thermal and ionization state of the intergalactic medium.  We investigate the robustness of such models of the \lya\ forest, focussing on the effect of particle initial conditions on the Lyman-$\alpha$ forest statistics in cosmological SPH simulations.  We study multiple particle initialization algorithms in simulations that are designed to be identical in other respects.  In agreement with the literature, we find that the correct linear theory evolution is obtained when a glass-like configuration is used for initial unperturbed gas particle positions alongside a regular grid configuration for dark matter particles and the use of non-identical initial density perturbations for gas and dark matter.  However, we report that this introduces a large scale-dependent distortion in the one-dimensional \lya\ transmission power spectrum at small scales ($k > 0.05$~s/km).  The effect is close to 50\% at $k\sim 0.1$~s/km, and persists at higher resolution.  This can severely bias inferences in parameters such as the dark matter particle mass.  By considering multiple initial conditions codes and their variations, we also study the impact of a variety of other assumptions and algorithmic choices, such as adaptive softening, background radiation density, particle staggering, and perturbation theory accuracy, on the matter power spectrum, the \lya\ flux power spectrum, and the \lya\ flux PDF.  This work reveals possible pathways towards more accurate theoretical models of the \lya\ forest to match the quality of upcoming measurements.
\end{abstract}


\section{Introduction}

The \lya\ forest absorption features seen in high-redshift quasar spectra are the pre-eminent tool for probing the small-scale structure of the Universe.  Properties of the absorption lines that constitute the \lya\ forest are sensitive to the small-scale matter power spectrum as well as the thermal evolution of baryons.  Furthermore, the forest is relatively insensitive to the non-linear baryonic physics at larger overdensities.  Consequently, hydrodynamical cosmological simulations can be used in combination with \lya\ forest measurements to put excellent constraints on various cosmological properties  (see \citealt{2009RvMP...81.1405M} and \citealt{2016ARA&A..54..313M} for reviews).

Examples of constraints on the cosmological small-scale structure include constraints derived by various authors from the \lya\ forest on dark matter models that affect the matter power spectrum at small scales.  Multiple authors have inferred lower bounds on the warm dark matter particle mass using \lya\ forest statistics \citep{2005PhRvD..71f3534V, 2008PhRvL.100d1304V, 2013PhRvD..88d3502V, 2017PhRvD..96b3522I, 2017JCAP...12..013B, 2020JCAP...04..038P, 2023PhRvD.108b3502V}, with recent work reporting $m_x>5.5$~keV \citep{2023arXiv230904533I}.  Similar work has yielded constraints on the mass of a sterile neutrino \citep{2005PhRvD..71f3534V, 2008PhRvL.100d1304V, 2017JCAP...12..013B, 2019MNRAS.489.3456G} and fuzzy or ultralight axion dark matter mass \citep{2017PhRvL.119c1302I, 2017MNRAS.471.4606A, 2020PhRvD.101l3518I, 2021PhRvL.126g1302R}.  Within the $\Lambda$CDM model, numerous authors have used the \lya\ forest to constrain the thermal state of the IGM \citep{2000MNRAS.318..817S, 2000ApJ...534...41R, 2001ApJ...557..519Z, 2002ApJ...567L.103T, 2010ApJ...718..199L, 2011MNRAS.410.1096B, 2015ApJ...799..196L, 2017PhLB..773..258G, 2017Sci...356..418R, 2017MNRAS.466.2690R, 2018MNRAS.474.2871R, 2019ApJ...872..101B, 2019ApJ...872...13W, 2019ApJ...876...71H, 2019ApJ...876...31G, 2021MNRAS.506.4389G, 2020MNRAS.494.5091G, 2022ApJ...933...59V}.  This has resulted in the broad picture that the IGM temperature $T_0$ at mean density rises from about 5,000~K at $z\gtrsim 5.4$, peaks at 15,000~K at $z\sim 3$, and reduces to 7,000~K at $z\lesssim 2$.

Most recent measurements of the statistics of the \lya\ forest have utilised measurements of a handful of quasars with spectrographs on 8m-class telescopes, such as Keck/HIRES, VLT/UVES, and VLT/XSHOOTER \citep{2018ApJ...852...22W, 2019ApJ...872..101B, 2022MNRAS.509.2842K}.  These measurements have signal-to-noise ratios of 10--30.  The typical resolution is medium-to-high, with some spectra having a resolution of $R\sim 4000$--$7000$, and some $R>40000$.  This allows a measurement of the one-dimensional \lya\ flux power spectrum to well above $k=0.1$~s/km, although at these small scales, contamination due to metal-line absorption could become a serious problem \citep{2018ApJ...852...22W}.  The resultant precision on the measurement of the flux power spectrum at $k\sim 0.1$~s/km is about 10\%.  At scales $k\lesssim 0.1$~s/km, the systematic errors are estimated to be about 20\% of the statistical errors \citep{2022MNRAS.509.2842K}.  Low-resolution quasar spectra from surveys like SDSS, BOSS, and eBOSS, containing tens to hundreds of thousands of samples, are also utilized for \lya\ forest studies. These spectra have limited resolution and signal-to-noise ratios. They are valuable for measuring neutrino masses and background cosmology parameters but cannot effectively constrain warm dark matter cosmologies or the thermal parameters of the IGM due to their insufficient resolution for measuring the \lya\ forest at small scales with $k>0.02$~s/km.

The quality of \lya\ forest data at small as well as large scales is expected to improve dramatically in the near future. DESI will increase the number of quasar spectra by an order of magnitude \citep{2016arXiv161100036D}, leading to order-of-magnitude improvement in constraints on the slope $n_\mathrm{s}$ and the running $\alpha_\mathrm{s}$ of the primordial curvature power spectrum, and the sum of neutrino masses, and a factor-of-two improvement in constraints on the effective number of neutrino species.  On the small scales, improvement is expected from projects such as WEAVE-QSO \citep{2016sf2a.conf..259P}.  If the large-scale measurements have a spectral resolution of $R\sim 2000$ and a signal-to-noise ratio of $\sim 5$, and the small-scale measurements have $R\sim 40000$ and a signal-to-noise ratio of $\gtrsim 30$, the WEAVE-QSO survey will substantially push the amount of intermediate quality data with $R\sim 20000$ and a signal-to-noise ratio of $\gtrsim 20$.  This will provide a means to study large-scale effects such as IGM temperature fluctuations.  A third improvement is the discovery of more high-redshift quasars by surveys such as LSST and Euclid.  In the last decade, the number of quasars with $z>6$ grew fivefold thanks to deep wide-field optical-infrared surveys such as CFHQS, VISTA/VIKING, Pan-STARRS1, DECaLS, UKIDDS, and UHS.  This pushed the number of high-redshift targets for \lya\ forest studies to 200.  The resultant increase in the number of \lya\ forest measurements has already allowed qualitatively new deductions about the high-redshift universe \citep{2020MNRAS.494.5091G, 2022MNRAS.514...55B}.  LSST will potentially increase this number to 10,000 \citep[for $i<26$;][]{2009arXiv0912.0201L}.  In the nearer term, Euclid is expected to find about 125 quasars with $7<z<8$, with potentially a handful of quasars with even higher redshifts \citep{2019A&A...631A..85E}.  Over a longer time scale, 30m-class telescopes will give a further boost to \lya\ forest data, increasing the quality and quantity of spectra to enable new ways of probing the IGM \citep{2015RAA....15.1945S, 2013arXiv1310.3163M}.
  
These forthcoming improvements to data make it imperative that theoretical modelling of the \lya\ forest matches measurements in accuracy.  Not only would such models be required to be free of numerical errors to a high degree, but they will also need to avoid poorly understood physical assumptions.  A large amount of recent work has focussed on this goal.  \citet{2022MNRAS.509.6119M} and \citet{2019MNRAS.490.3177W} investigated the effect of ignoring reionization-induced large-scale spatial fluctuations in the IGM temperature on the one-dimensional flux power spectrum of the \lya\ forest.  \citet{2022MNRAS.509.6119M} found that for the 10\% uncertainties on current measurements, temperature fluctuations do not introduce any biases in the inference of dark matter or thermal parameters.  However, for 5\% uncertainties, such as those expected in future data, they find systematic shifts of about 1$\sigma$.  \citet{2022arXiv220705023C} investigated the difference between results from smoothed particle hydrodynamics (SPH) and Eulerian hydrodynamical computation.  They found the one-dimensional power spectrum results from both techniques agreed to be better than a percent.  They also found that SPH leads to systematic biases in several \lya\ forest statistics at high redshift because of poor sampling in low-density regions.  \citet{2009MNRAS.398L..26B} investigated the effect of mass resolution and box size of SPH simulation on the \lya\ forest spectra for $2<z<5$.  They found that the requirements on simulations become more stringent towards higher redshifts because the typical densities probed by the \lya\ absorption lines are relatively lower.  \citet{2021JCAP...04..059W} did a similar analysis for Eulerian codes, finding that a box size of 120~Mpc and a resolution of 30~kpc is required for one-dimensional flux power spectrum predictions that are accurate to 1\%.  These authors also investigated the accuracy of the interpolation required to span the parameter space to accurately infer model parameters from the data. \citet{2013MNRAS.429.3341B} also compared Eulerian and SPH methods for modelling the \lya\ forest.  Although they found some differences between the two models at large neutral-hydrogen column densities, there were no significant differences in the column densities probed by the \lya\ forest.  \citet{2020JCAP...06..002B} showed that most simulations of the \lya\ forest produce a large error in the linear-theory predictions for perturbations in dark matter and baryon densities.  They argued that this affects the one-dimensional flux power spectrum predictions at the 5\% level.  \citet{2021ApJ...912..138V} investigated the role of photoheating and photoionization models in \lya\ forest simulations using their Eulerian code.

In this paper, we turn our attention to the accuracy of the initial conditions used in cosmological hydrodynamical simulations of the \lya\ forest.  In recent times, several different algorithms and codes for generating initial conditions have been used, without a detailed understanding of the robustness of each of them for \lya\ forest statistics.  We set up identical simulations using multiple codes and compare multiple forest statistics.  The outline of this paper is as follows.  Section~\ref{sec:methods} describes the current landscape of methods used to set up initial conditions.  We describe our set-up in Section~\ref{sec:setup}.  Our results are presented in Section~\ref{sec:results}.  We end with a discussion of our findings in Section~\ref{sec:discuss} and a summary in Section~\ref{sec:conclusions}.  Our $\Lambda$CDM cosmological parameters are  $\Omega_\mathrm{b}=0.0482$, $\Omega_\mathrm{m}=0.308$, $\Omega_\Lambda=0.692$, $h=0.678$, $n_\mathrm{s}=0.961$, $\sigma_8=0.829$, and $Y_\mathrm{He}=0.24$ \citep{2014A&A...571A..16P}.

\section{Initial Conditions in SPH-based Cosmological Simulations}
\label{sec:methods}

The task of setting up initial conditions for cosmological simulations is to obtain a particle distribution inside a finite volume that represents a density field with Gaussian random fluctuations around the mean.  Such a density field should model cosmological density perturbations at sufficiently early times in the matter-dominated era.  The resultant density contrast
\begin{equation}
  \delta(\mathbf{x}, t) = \frac{\rho(\mathbf{x}, t)}{\bar\rho(t)} - 1 ,
\end{equation}
can be expanded into its Fourier components as 
\begin{equation}
  \delta(\mathbf{x}, t) = \frac{1}{(2\pi)^{3/2} }\int_{-\infty}^\infty \ud^3k\,\tilde\delta(\mathbf{k}, t) \exp{(i \mathbf{k} \cdot \mathbf{x})}.
  \label{eqn:dk}
\end{equation}
As $\delta$ is a Gaussian random field with zero mean, it is fully specified by its power spectrum, $P(k, t)$, defined by 
\begin{equation}
  \langle\tilde\delta(\mathbf{k}, t)\,\tilde\delta^*(\mathbf{k^\prime}, t)\rangle = P(k,t)\,\delta_\mathrm{D}(\mathbf{k}-\mathbf{k^\prime}),
  \label{eqn:pk}
\end{equation}
where the angle brackets denote an ensemble average, $k$ is the magnitude of the vector $\mathbf{k}$, and $\delta_\mathrm{D}$ is the Dirac delta function.  The power spectrum $P(k)$ has units of volume, often chosen to be $(\mathrm{cMpc}/h)^3$.  When perturbations are small, the power spectrum is described by a transfer function $T(k,t)$, so that
\begin{equation}
  P(k, t) = T^2(k) D^2(t) P_\mathcal{R}(k),
  \label{eqn:tk}
\end{equation}
where $D(t)$ is the linear growth factor, and $P_\mathcal{R}(k)$ is the dimensionless primordial curvature power spectrum, taken to be of the form
\begin{equation}
  P_\mathcal{R}(k) = A_\mathrm{s}\left(\frac{k}{k_0}\right)^{n_\mathrm{s}},
\end{equation}
where, with $k_0=0.05$~cMpc$^{-1}$, measurements favour the power-law index $n_\mathrm{s}=0.961$ and $A_\mathrm{s}=2.2\times 10^{-9}$ \citep{2014A&A...571A..16P}.

In practice, one has to work with quantities defined on a set of discrete points.  Consider a Cartesian grid $\mathbf{x}_p=p\,\Delta x$, with a uniform spacing $\Delta x$ in the three directions, spanning the simulation volume.   We assume that there are $n$ points in each direction, for a total of $n^3$ points.  When the density contrast is evaluated at these discrete points, we can compute the discrete Fourier transform as
\begin{equation}
  \tilde\delta_{\mathbf{k}_q} = \frac{1}{n^{3/2}}\sum_{p=0}^{n-1}\delta(\mathbf{x}_p)\exp{(-i\mathbf{k}_q\cdot\mathbf{x}_p)},
\end{equation}
where $\mathbf{k}_q=2\pi q/(n\,\Delta x)$ for $q=0, \ldots, n-1$.  The quantity $\tilde\delta_{\mathbf{k}_q}$ is related to the quantity $\tilde\delta(\mathbf{k})$ defined in Equation~(\ref{eqn:dk}) by
\begin{equation}
  \tilde\delta(\mathbf{k}_q) = (\Delta x)^3\left(\frac{n}{2\pi}\right)^{3/2}\tilde\delta_{\mathbf{k}_q}.
\end{equation}
This allows us to compute the power spectrum from Equation~(\ref{eqn:pk}) at discrete points $k_q$ as
\begin{equation}
  P(k_q)=(\Delta x)^3\langle|\tilde\delta_{\mathbf{k}_q}|^2\rangle,
  \label{eqn:pk_discrete}
\end{equation}
where the angle brackets now denote the mean over all grid points. All three-dimensional power spectra in this paper are computed using Equation~(\ref{eqn:pk_discrete}).

To specify the initial conditions of a cosmological simulation, one must compute the transfer function $T(k)$, defined by Equation~(\ref{eqn:tk}).  This can be done to 0.1\% accuracy using Boltzmann solver codes such as CAMB \citep{2000ApJ...538..473L} or CLASS \citep{2011arXiv1104.2932L}.

\begin{table*}
  \begin{tabular}{llll}
    \hline
    Code & Perturbation sampling & Displacement computation & Velocity computation \\
    \hline 
    \ngenic & Fourier space & Zel'dovich & Equations~(\ref{eqn:v1}) and (\ref{eqn:v2}) \\
    \tlpt & Fourier space & 2LPT & Equation~(\ref{eqn:v_2lpt}) \\ 
    \cosmic & real space & Zel'dovich & Equations~(\ref{eqn:v1}) and (\ref{eqn:v2}) \\ 
    \mpgenic & Fourier space & Zel'dovich & Equations~(\ref{eqn:v3}) and (\ref{eqn:v4}) \\ 
    \music & real space & Zel'dovich or 2LPT & Equations~(\ref{eqn:v1}) and (\ref{eqn:v2}) \\
    \hline
  \end{tabular}
  \caption{An overview of important methodological similarities and differences between the initial conditions codes investigated in this paper. Although \ngenic, \cosmic, and \music\ use the same Equations~(\ref{eqn:v1}) and (\ref{eqn:v2}) for computing velocities, there are small differences in their implementation. \ngenic\ approximates the quantity $F(z)$ by $\Omega_\mathrm{m}^{0.6}(z)$. \cosmic\ and \music\ obtain $F(z)$ using linear perturbation theory, however \music\ includes the contribution of the radiation energy density while doing so, whereas \cosmic\ does not.} See text for details. \label{tab:code}
\end{table*}

\subsection{Generating initial conditions}

With the above notation in place, we can now describe the general procedure of setting up initial conditions for a cosmological simulation.  This task can be split into two steps.  The first step involves creating a uniform distribution of particles.  The second step is to displace these particles to achieve a Gaussian random density distribution that has the required power spectrum.

It is not trivial to set up a uniform mass distribution with a finite number of particles \citep{1998ARA&A..36..599B}.  Simply distributing $n$ particles randomly in the box leads to a density distribution with a non-zero white-noise power spectrum that can lead to non-trivial structure formation even in the absence of additional cosmological power.  A conventional solution to this problem is to position the particles in a regular Cartesian cubic grid.  This method is widely used in the literature, although it does suffer from problems.  For instance, the grid spacing imposes a characteristic length scale in the simulation.  The grid also breaks isotropy at small scales by defining preferred directions.  These problems are particularly important to address in cosmological models with little or no power at small scales, such as warm-dark-matter cosmologies.  An alternative to using a grid distribution is to create a so-called glass-like distribution of particles \citep{1995MNRAS.274.1049B, 1996clss.conf..349W}.  To construct a glass-like distribution, one starts from a random distribution of particle positions, which is then evolved using a repulsive gravitational force.  The glass then shows no visible order or anisotropy on scales beyond a few interparticle separations.

Once a uniform particle distribution is obtained, initial conditions for cosmological simulations are derived by shifting these particle positions by displacements designed to get a Gaussian random density distribution with the desired power spectrum.  This is achieved in several stages.  First, random numbers representing Gaussian random density perturbations with a flat, white-noise power spectrum are sampled using a random number generator, in real or Fourier space, and then transformed so that they have the desired power spectrum \citep{1985MNRAS.217..805P, 1986ApJ...304...15B, 1996ApJ...460...59S}.  Next, particle displacements are computed from these sampled overdensities.  This can be done by mapping the overdensities to gravitational potential perturbations, which then yield the concomitant particle displacements.  The displacements are used to modify the uniform particle positions of the initial glass or grid. Lastly, consistent velocities are assigned to the particles.

While this outline of the procedure to generate initial conditions for cosmological simulations is generally uniform across codes, many algorithmic variations exist.  Some codes use particle grids while others use particle glasses.  Some codes stagger gas and dark matter particles by adding a small initial separation between particle positions, but leave this separation to the user's choice.  Some codes sample white-noise density perturbations in real space while others sample them in Fourier space.  The computation of particle displacements from the density perturbations is done with a variety of assumptions and accuracy levels of perturbation theory by different codes.  Our purpose in this paper is to understand how these differences in initial condition codes affect cosmological simulation results relevant to the \lya\ forest.  In order to do this, we now describe the methods behind the initial conditions codes that we compare in this paper, before looking at the results from cosmological simulations using these codes.

\subsection{Initial conditions codes}

We compare five initial conditions codes in this paper: \ngenic, \tlpt, \cosmic, \mpgenic, and \music.  We begin by giving a brief description of the algorithms of these codes.

\subsubsection{\ngenic}
\label{sec:ngenic}

\ngenic\ \citep{2005Natur.435..629S, 2012MNRAS.426.2046A} allows the use of either glass and grid particle distributions, and uses the \citet{1970A&A.....5...84Z} approximation to compute the displacements.  Gaussian density perturbations with the desired power spectrum are sampled in \ngenic\ in Fourier space.  This is done by drawing random samples $x$ and $\phi$ from uniform distributions between 0 and 1, and 0 and $2\pi$, respectively.  Then, $\tilde\delta_{\mathbf{k}}=Re^{i\phi}$ are the required Fourier-space density perturbations with power spectrum $P(k)$ if $R=(\Delta k)^{3/2} \sqrt{-P(k)\log{x}}$ with $(\Delta k)^3$ being the Fourier-space volume element defined by Equation~(\ref{eqn:pk_discrete}).  The Hermitian symmetry condition, $\tilde\delta_{\mathbf{k}}^*=\tilde\delta_{\mathbf{-k}}$, is imposed to ensure that the corresponding real-space perturbations are real.

Using the density perturbations, particle displacements $\mathbf{d}_\mathbf{k}$ are computed using the Zel'dovich approximation as \citep{1970A&A.....5...84Z}
\begin{equation}
  \mathbf{d}_{\mathbf{k}} = \frac{i\mathbf{k}}{k^2}\delta_{\mathbf{k}}.
  \label{eqn:za}
\end{equation}
These displacements are then transformed to the real space.  When the initial uniform particle distribution is on a grid, one can simply add the real-space displacements to the particle positions to get the desired particle distribution.  This step can be less than straightforward when the initial uniform particle distribution is not on a grid.  In this case, \ngenic\ obtains the displacements at the particle positions by trilinear interpolation on the displacement grid.

Finally, velocities are assigned to the displaced particles by using the equations of motion to write 
\begin{equation}
  \mathbf{v}(\mathbf{x}) = a H(z) F (z) \mathbf{d}(\mathbf{x}),
  \label{eqn:v1}
\end{equation}
where $\mathbf{d}(\mathbf{x})$ are the real-space particle displacements, $H$ is the Hubble constant, $z$ is the redshift at which the initial conditions are desired, and
\begin{equation}
  F(z) = - \frac{\text{d}\ln{D(z)}}{\text{d}\ln{(1+z)}}.
  \label{eqn:v2}
\end{equation}
In \ngenic, $F(z)$ is approximated to be $\Omega^{0.6}_\mathrm{m}(z)$, where $\Omega_\mathrm{m}=\rho_\mathrm{m}(z)/\bar\rho(z)$ is the cosmic matter density relative to the mean. 
This grid-based \ngenic\ set-up was used in the simulations belonging to the Sherwood \citep{2017MNRAS.464..897B} and the Sherwood-Relics \citep{2023MNRAS.519.6162P} suites.   

\subsubsection{\tlpt}

\tlpt\ \citep{1998MNRAS.299.1097S, 2006MNRAS.373..369C, 2012PhRvD..85h3002S} is closely related to \ngenic.  The two codes follow identical procedures for setting up the initial uniform particle distributions and density perturbations.  They differ in the computation of displacements and velocities, however, as \tlpt\ uses second-order Lagrangian perturbation theory instead of the Zel'dovich approximation discussed above.  In this way, the particle displacements in Fourier space are given by
\begin{equation}
  \mathbf{d}_{\mathbf{k}} = \mathbf{d}^{(1)}_{\mathbf{k}} + \mathbf{d}^{(2)}_{\mathbf{k}},
\end{equation}
where $\mathbf{d}^{(1)}_{\mathbf{k}}$ is the first-order contribution to the displacement, given by the Zel'dovich approximation via Equation~(\ref{eqn:za}), and $\mathbf{d}^{(2)}_{\mathbf{k}}$ is the second-order contribution given by 
\begin{equation}
  \mathbf{d}^{(2)}_{\mathbf{k}} = - \frac{3}{7}\frac{\textbf{k}\Tilde{S}_{\mathbf{k}}}{ik^2},
\end{equation}
where $\Tilde{S}_{\mathbf{k}}$ is the discrete Fourier transform of
\begin{equation}
  S = \sum_{i > j} \left[d^{(1)}_{i,i} d^{(1)}_{j,j} -  (d^{(1)}_{i,j})^2\right],
\end{equation}
with $d^{(1)}_{i,j}$ being the derivative of the $i$th component of the first-order real-space displacement vector $\mathbf{d}$ with respect to the $j$th component of the position vector,
\begin{equation}
  d^{(1)}_{i,j} = \frac{\partial d^{(1)}_i}{\partial x_j} = ik_jd^{(1)}_i.
\end{equation}
The corresponding velocities are given by
\begin{equation}
  \mathbf{v}(\mathbf{x}) = aH(z) \Omega^{3/5}_\mathrm{m}(z)\mathbf{d}^{(1)}(\mathbf{x}) + 2 aH(z)\Omega^{4/7}_\mathrm{m}(z)\mathbf{d}^{(2)}(\mathbf{x}).
  \label{eqn:v_2lpt}
\end{equation}

\subsubsection{\cosmic}

\cosmic\ \citep{2015MNRAS.446.3697L} is a Zel'dovich code that uses a grid for the initial particle distribution.  The code does not itself separate gas and dark matter particles, it instead leaves this task to the gravity and hydrodynamics solver.  To obtain the density field, \cosmic\ samples density values $\delta(\mathbf{x}_p)$ from a Gaussian distribution with zero mean and unit variance at each point on a grid in the real space.  These white-noise perturbations are then discrete-Fourier transformed to Fourier space.  The real and imaginary parts of these Fourier-transformed perturbations then follow a Gaussian distribution with zero mean and variance $\sqrt{n/2}$.  These perturbations are then multiplied by $(\Delta k)^{3/2}\sqrt{P(k)/n}$ to obtain Gaussian density perturbations in the Fourier space with the given power spectrum $P(k)$.  Displacements are then computed from these Fourier space density perturbations using the Zel'dovich approximation of Equation~(\ref{eqn:za}).  The corresponding velocities are computed using Equations~(\ref{eqn:v1}) and (\ref{eqn:v2}), however unlike \ngenic, \cosmic\ does not approximate the quantity $F(z)$ with $\Omega_\mathrm{m}^{0.6}(z)$. Instead, \cosmic\ computes $F(z)$ using linear perturbation theory. While doing so it ignores the radiation energy density.

\subsubsection{\mpgenic}

\mpgenic\ \citep{2020JCAP...06..002B} allows for various ways of initialising the particle positions.  \citet{2020JCAP...06..002B} showed that initialising dark matter particles on a grid and gas particles on a glass reproduces the expected linear evolution of the difference between the power spectra of gas and dark matter.  This is the strategy we adopt while using \mpgenic\ in this paper. \mpgenic\ generates Gaussian density perturbations with a given density power spectrum in an identical manner to \ngenic\ and \tlpt.  The gas and dark matter densities are then obtained by scaling the total matter density perturbations with the Fourier-space gas and dark matter density fractions obtained from CLASS \citep{2011arXiv1104.2932L}.  The corresponding displacements are obtained by means of the Zel'dovich approximation, via Equation~(\ref{eqn:za}).  This displacement field is then transformed to real space.  Gas particle displacements are then obtained using trilinear interpolation.  Velocities are obtained from CLASS as transfer functions 
\begin{equation}
  v_{\text{CDM}}(k) = \dot\delta_{\text{CDM}}(k) = - \frac{\dot{h}(k)}{2},
  \label{eqn:v3}
\end{equation}
and
\begin{equation}
  v_{\text{b}}(k) = \dot\delta_{\text{b}}(k) = -\theta_{\text{b}}(k) - \frac{\dot{h}(k)}{2},
  \label{eqn:v4}
\end{equation}
where $h$ is the synchronous gauge density perturbation, and $\theta_{\text{b}}$ is the baryon velocity divergence, both computed by CLASS. These velocity transfer functions are sampled similar to the density transfer functions.  The resultant Fourier-space velocities are then transformed to real space and interpolated using CIC.  \mpgenic\ avoids spurious gravitational coupling between gas and dark matter particles by evolving the pre-displaced combined distribution of the two types of particles with reversed gravity for a small number of time steps. 

\begin{figure}
  \begin{center}
    \includegraphics*[width=\columnwidth]{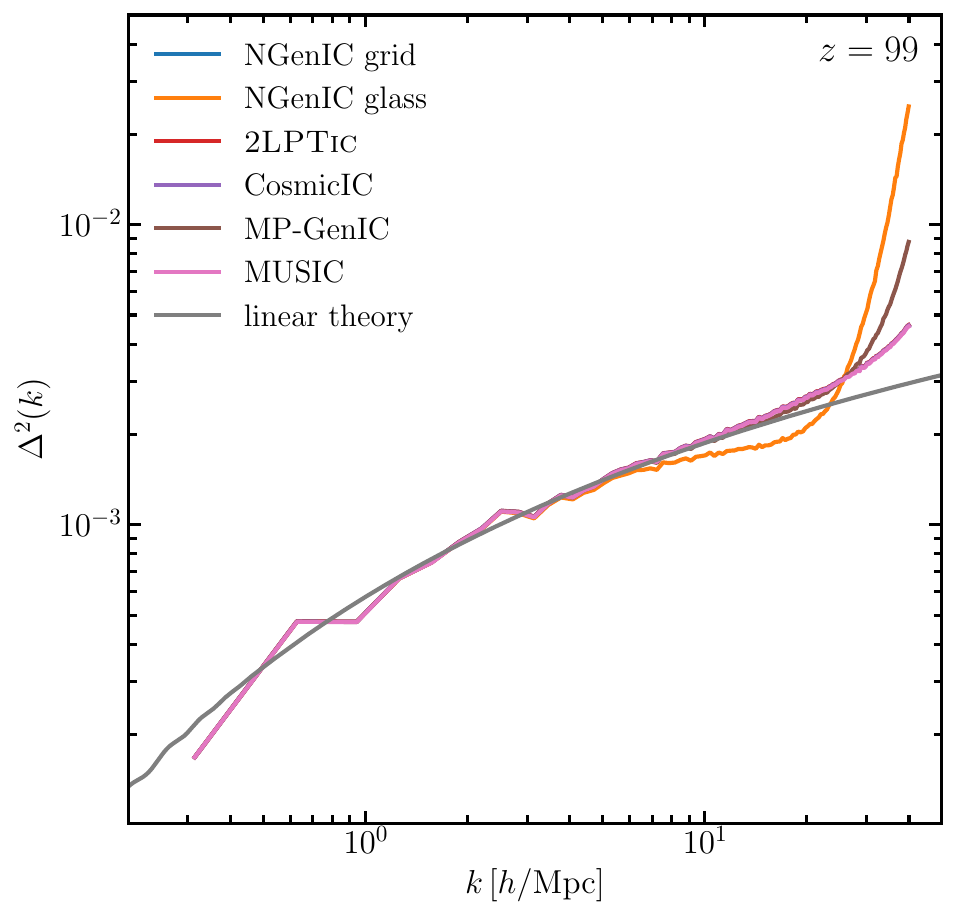}
  \end{center}
  \caption{The three-dimensional power spectrum of the total matter density in the initial conditions used for our simulations, at $z = 99$.  We show results from the five codes that we study.  For \ngenic, we also show the result when using a glass-like instead of a grid-based particle distribution for the initial unperturbed matter density.  Most of the curves overlap each other for most of the range of $k$ values.  The grey curve is the linear theory power spectrum computed using CLASS \citep{2011arXiv1104.2932L}.  We see that in the initial conditions, the matter density power spectrum between the codes agrees very well.  The only significant outliers are the glass-based \ngenic\ and \mpgenic\ set-ups.}
  \label{fig:InitialPS}
\end{figure}

\subsubsection{\music}

\music\ \citep{music} allows initialising the particle load using either a grid or a glass.  We use the grid in this work.  In contrast with \ngenic, \music\ obtains real-space Gaussian density perturbations with a given power spectrum by sampling a Gaussian variate with zero mean and unit variance on a grid.  This Gaussian variate is then convolved with the inverse Fourier transform of $\sqrt{P(k)}$ to get the desired density perturbations $\delta(\mathbf{x})$ on the grid.  Particle displacements are then obtained in a similar way as \ngenic\ following Equation~(\ref{eqn:za}).  Velocities are also computed similar to \ngenic, using Equations~(\ref{eqn:v1}) and (\ref{eqn:v2}). However, unlike \ngenic, \music\ computes the quantity $F(z)$ using linear perturbation theory. While doing so, unlike \cosmic, \music\ includes the radiation energy density. Finally, corrections are applied to these velocities to compensate for inaccuracies caused by discreteness.

\subsection{Initial conditions for different particle initialisation choices}

Our fiducial simulations consist of five runs with the five initial condition set-ups from Table~\ref{tab:code}.  Figure~\ref{fig:InitialPS} shows the three-dimensional spherically-averaged power spectrum of the total matter density contrast at $z=99$ from our five fiducial simulations.  Also shown in this figure is the linear theory result at this redshift, obtained using CLASS \citep{2011arXiv1104.2932L}.  The figure also shows the power spectrum from the glass-based \ngenic\ run, to compare to the corresponding grid-based run.

We see that all grid-based simulations reproduce the theoretical linear power spectrum extremely well.  (Most curves in Figure~\ref{fig:InitialPS} completely overlap with each other.) The only case with some difference is \tlpt, where the use of the second-order perturbation theory results in differences from our fiducial run, but even these differences are sub-percent.  In contrast to these results, a large difference is seen in the glass-based \ngenic\ run, which deviates from its grid-based counterpart with suppression in power of close to 20\% at $k=20\;h/$cMpc and a strong enhancement in power of up to a factor of 2 at larger values of $k$.  This excess power in the glass-based set-up is the well-known $k^4$ enhancement due to the growth of random gravitational instabilities during the creation of the glass configuration \citep{2020JCAP...06..002B}.  Indeed, a similar enhancement is also seen in the \mpgenic\ case, although this is relatively smaller because \mpgenic\ uses glass only for the baryons.

\begin{figure}
  \begin{center}
    \includegraphics*[width=\columnwidth]{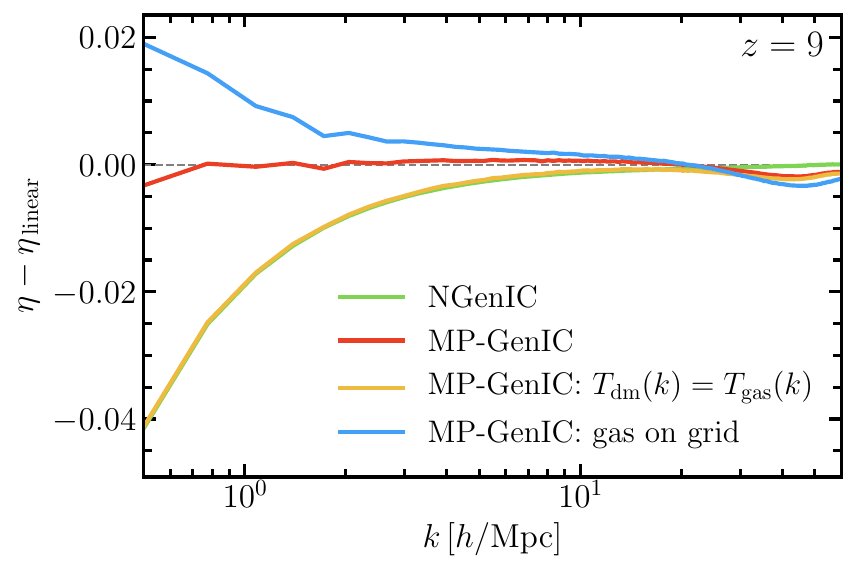}
  \end{center}
  \caption{The difference between the linear theory value of $\eta=(\tilde\delta_\mathrm{dm}-\tilde\delta_\mathrm{gas})/2$ and its value in four of our simulations at $z=9$.  The green and red curves show the power difference from our fiducial grid-based \ngenic\ run and our fiducial \mpgenic\ run, respectively.  The yellow curve shows the result when we use identical transfer functions for gas and dark matter in the \mpgenic\ run, while the blue curve shows the same when we use a grid-based initialisation for gas particles in the \mpgenic\ run.  As previously found by \citet{2020JCAP...06..002B}, we see that only fiducial \mpgenic\ simulation recovers the correct linear theory evolution.} \label{fig:Bird_Fig5}
\end{figure}

While such small-scale enhancement suggests a numerical error, \citet{2020JCAP...06..002B} pointed out that there might be advantages in initialising gas particles on a glass.  They argued that the \mpgenic\ set-up, which uses distinct transfer functions for gas and dark matter, and uses a glass-like configuration for gas particles, can reproduce the linear theory evolution of the power difference between gas and dark matter.  Figure~\ref{fig:Bird_Fig5} shows this at $z=9$.  We show the difference in the dark matter and gas power spectra in four runs, by plotting the quantity $\eta=(\tilde\delta_\mathrm{dm}-\tilde\delta_\mathrm{gas})/2$ at $z=9$.  (The simulation details on the evolution from $z=99$ to $z=9$ are described in the next section.)  The green curve shows the power difference from our fiducial grid-based \ngenic\ run, and the red curve shows the same quantity for our fiducial \mpgenic\ run.  The yellow curve shows how the result changes when we use identical transfer functions for gas and dark matter in the \mpgenic\ run.  The blue curve shows the power difference when we use a grid-based initialisation for gas particles in the \mpgenic\ run. For all four cases we have shown the difference between simulated and the linear theory value of $\eta$. In the linear theory, the power in dark matter and baryons is calculated from the CLASS output using the appropriate columns in the transfer function and total matter power spectrum files. The transfer functions are calculated as $\delta_\mathrm{cdm/b} = \delta_\mathrm{cdm/b} / \delta_\mathrm{total}$, where $\delta_\mathrm{cdm/b}$ are from the CLASS outputs and $\delta_\mathrm{total} = (\Omega_\mathrm{b} \delta_\mathrm{b} + \Omega_\mathrm{cdm} \delta_\mathrm{cdm}) / (\Omega_\mathrm{b} + \Omega_\mathrm{cdm})$. The powers for both species are given by $\delta_\mathrm{cdm/b} = \delta_\mathrm{cdm/b} \sqrt{P_\mathrm{tot}}$ and $P_\mathrm{tot}$ is obtained from the CLASS power spectrum outputs.  As noted by \citet{2020JCAP...06..002B}, we see here that only the fiducial \mpgenic\ run manages to reproduce the expected linear evolution.  Removing the glass, or removing the difference between the gas and dark matter transfer functions results in a significant disagreement with the linear theory.

We will now see how these particle initialisation choices affect the small-scale structure in the \lya\ forest.

\section{Non-linear Evolution}
\label{sec:setup}

Our procedure for computing the non-linear gravitational and hydrodynamical evolution is identical to the set-up used in the Sherwood Simulation Suite \citep{2017MNRAS.464..897B}.  We perform our simulations using the SPH code \gadget, the well-known modified version of the publicly available Gadget2 code \citep{2005Natur.435..629S}. In this code the gravitational forces are computed using the TreePM algorithm where short-range forces are obtained using a tree method and long-range forces are computed using mesh-based Fourier methods. In the tree algorithm, particles are grouped together according to their distances and the force from each group is approximated using a multipole expansion. In the particle-mesh (PM) algorithm the density field is realised on a mesh grid using CIC interpolation and the gravitational potential is constructed by solving the Poisson's equation.  The mesh dimension in each direction is set by the \texttt{PMGRID} parameter, which we set to the cube root of the particle number in all simulations presented in this paper.  To avoid a singularity in the force calculation when the particles move close to each other, a softening parameter $l_\mathrm{soft}$ is introduced.  We set this to 1/25th of the mean inter-particle separation.  Apart from \mpgenic, we use \gadget\ for splitting particles into gas and dark matter.  The dark matter and gas positions are then calculated by adding constant vectors to the total matter positions, keeping the centre of mass at the position of the original particle. The constant separations are
\begin{equation*}
	\delta_{\mathrm{gas}} = \frac{1}{2} \frac{\Omega_\mathrm{m} - \Omega_\mathrm{b}}{\Omega_\mathrm{m}} d,\quad\mathrm{and}
\end{equation*} 
\begin{equation*}
	\delta_{\mathrm{dm}} = \frac{1}{2} \frac{\Omega_\mathrm{b}}{\Omega_\mathrm{m}} d,
\end{equation*} 
where $d = L / N$ is the inter-particle spacing. Our fiducial simulation, described below, has $d = 78.125$\,ckpc$/h$, $\delta_{\mathrm{gas}} \sim 33$\,ckpc$/h$ and $\delta_{\mathrm{dm}} \sim 6$\,ckpc$/h$. These separations are added to the total matter positions along all three directions, so that $x_{\mathrm{gas}} = x_{\mathrm{gas}} - \delta_{\mathrm{gas}}$, $y_{\mathrm{gas}} = y_{\mathrm{gas}} - \delta_{\mathrm{gas}}$, and $z_{\mathrm{gas}} = z_{\mathrm{gas}} - \delta_{\mathrm{gas}}$ for the gas particles, and $x_{\mathrm{dm}} = x_{\mathrm{dm}} + \delta_{\mathrm{dm}}$, $y_{\mathrm{dm}} = y_{\mathrm{dm}} + \delta_{\mathrm{dm}}$, and $z_{\mathrm{dm}} = z_{\mathrm{dm}} + \delta_{\mathrm{dm}}$ for the dark matter particles.  Both gas and dark matter particle velocities are assigned to be the same as that of the original total matter particle.  In the case of the \mpgenic, the particles are separately supplied to \gadget\ as part of the initial conditions.

\begin{figure*}
  \begin{center}
    \includegraphics*[width=0.8\textwidth]{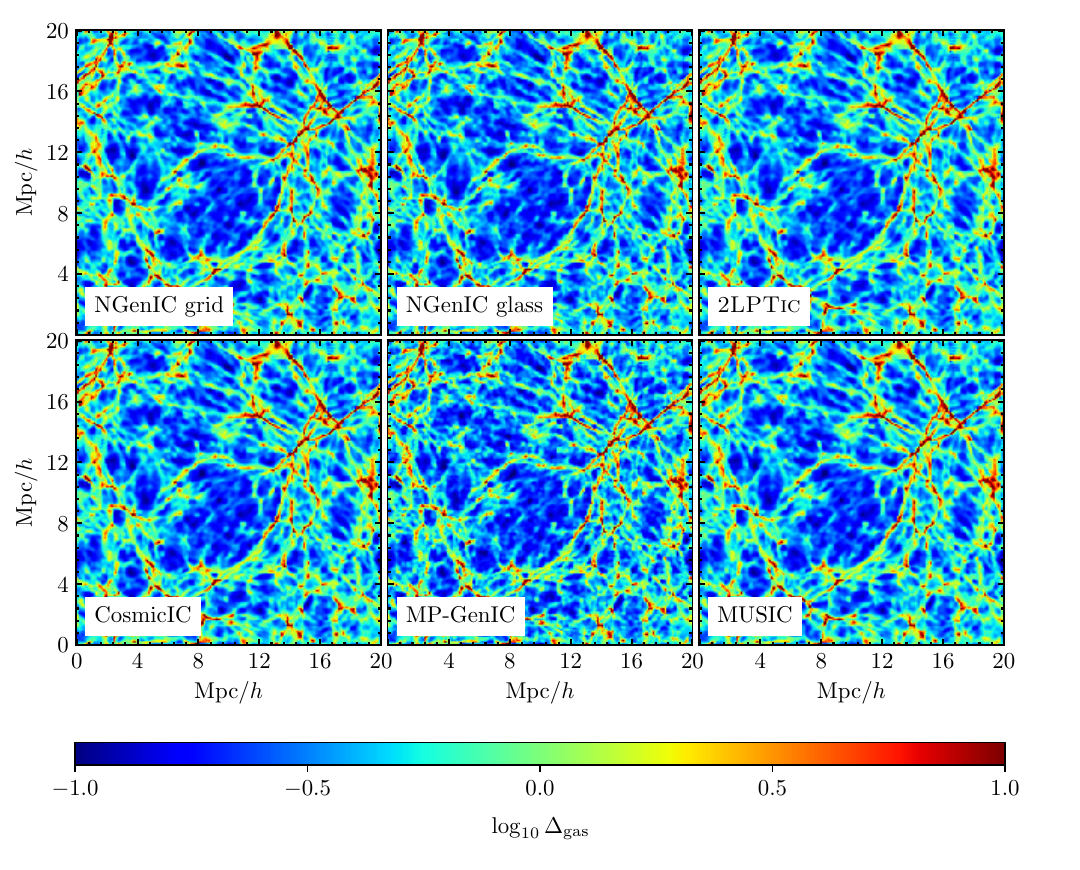}
  \end{center}
  \caption{The gas density distribution at $z=3$ for our six primary simulations. Each slice shows the gas density projected over 78.125~ckpc$/h$. The simulations produce similar gas distributions, thanks to our use of the identical random phases in all of them.
  \label{fig:slices}}
\end{figure*}

\begin{figure*}
  \begin{center}
    \includegraphics*[width=\textwidth]{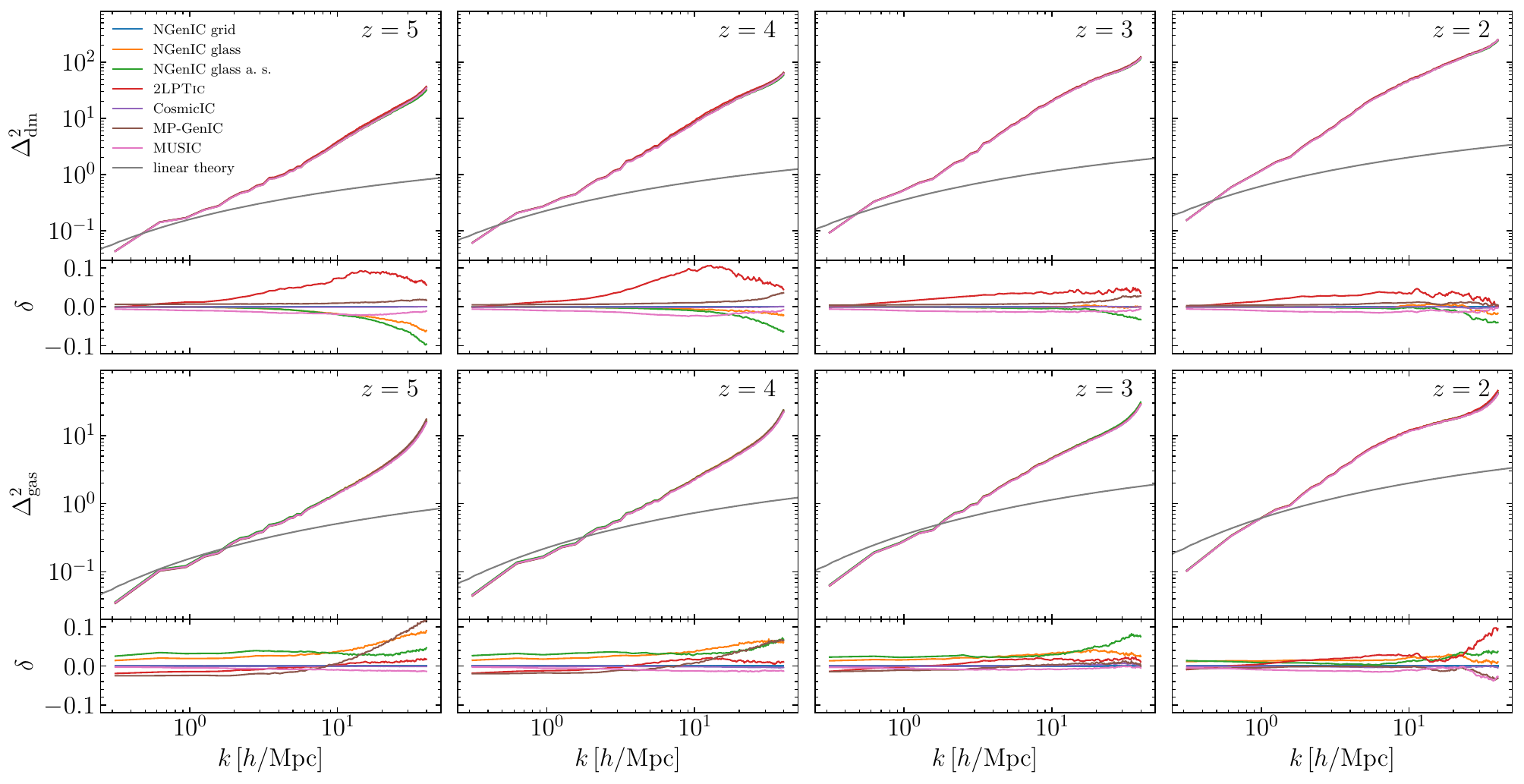}
  \end{center}
  \caption{The three-dimensional dimensionless power spectrum of the dark matter (top row) and gas density (bottom row) at $z=2,3,4,$ and $5$. Each panel shows results from our five primary simulations.  Also shown are results from a glass-based \ngenic\ run, and a glass-based \ngenic\ run that uses adaptive softening (indicated by `a.~s.' in the legend).  The linear theory power spectrum is also shown at each redshift.  The difference panels show the relative difference from the grid-based \ngenic\ results.}
  \label{dm_gas_ps}
\end{figure*}

We run our simulations down to $z=2$ and examine results at $z=9, 5, 4, 3,$ and $2$.  In addition to the cosmological evolution of baryons and dark matter, \gadget\ implements photoionization, photoheating, radiative cooling. We use the \texttt{QUICK\_LYALPHA} flag in the code to speed up the simulation by making gas particles with overdensities of greater than $10^3$ with temperature less than $10^5$~K collisionless and removing them from the hydrodynamic calculation.  This does not affect our study of the low-density IGM.  The metagalactic UV background is modelled according to a modified version of the \citet{2012ApJ...746..125H} reionization model (see \citealt{2017MNRAS.464..897B} for details).

Our fiducial simulations use $N = 2 \times 256^3$ particles to present the gas and the dark matter in a computational box of comoving length $20\,h^{-1}\text{Mpc}$. We refer to these as the 20--256 runs.  Periodic boundary conditions are implemented. The mass of dark matter particle is $M_{\text{dm}} = 3.44 \times 10^7 \, {\text{M}}_{\odot}/h$ and the gas particles are of mass $M_{\text{gas}} = 6.38 \times 10^6 \, {\text{M}}_{\odot}/h$.  We also study higher resolution 20--512 and 20--1024 simulations.  As their names suggest, these simulations use $2 \times 512^3$ and $2 \times 1024^3$ particles in the same volume as the 20--256 simulations.  In the 20--512 simulations, the dark matter and gas masses are $M_{\text{dm}} = 1.72 \times 10^7 \, {\text{M}}_{\odot}/h$, and $M_{\text{gas}} = 3.19 \times 10^6 \, {\text{M}}_{\odot}/h$, and we use $\texttt{PMGRID} = 512$ and $l_\mathrm{soft} = 1.5625 \, \text{kpc}$. For the 20--1024 simulations, these quantities take the values $M_{\text{dm}} = 0.86 \times 10^7 \, {\text{M}}_{\odot}/h$, $M_{\text{gas}} = 1.595 \times 10^6 \, {\text{M}}_{\odot}/h$, $\texttt{PMGRID} = 1024$ and $l_\mathrm{soft} = 0.78125 \, \text{kpc}$.

In order to compute the dark matter and gas density power spectra, we project the output of the hydrodynamical simulations on a uniform Cartesian grid.  The densities for the dark matter are calculated on a regular grid from the particle positions using the CIC kernel \citep{1981csup.book.....H}. The gas particles are projected on the grids by integrating the SPH kernel on grid cells. The density contrast $\delta = \rho / \Bar{\rho} -1 $ is calculated for both dark matter and gas particles on each grid. Then the power spectrum is then calculated using Equation~(\ref{eqn:pk_discrete}).  We set the number of cells in each direction for the grid in $k$-space to the cube root of the number of particles.  While computing the power spectrum of the dark matter density, we deconvolve the CIC kernel.

The \lya\ forest sightlines are extracted from the simulation box by selecting random points on the $xy$ plane, and choosing all lines to go in the $z$ direction. Then the Ly$\alpha$ optical depth is calculated as 
    \begin{equation}
        \tau (v) = \frac{3A_{2p1s}\lambda^3_\mathrm{Ly\alpha}}{8\pi H(z) } \int_{-\infty}^{\infty} n_\mathrm{HI} \phi(v) \text{d}v, 
    \end{equation}
    where $n_\mathrm{HI}$ is the neutral hydrogen number density, $\lambda_\mathrm{Ly\alpha}$ is the rest-frame Ly$\alpha$ wavelength, $A_{2p1s}$ is the Einstein $A$ coefficient and $\phi (v)$ is the Voigt line profile. The Ly$\alpha$ transmission is then $F = \exp{(-\tau)}$  Following standard practice for such optically thin simulations, we normalise the transmission such that the mean transmission has the observed value.

    While deriving the one-dimensional power spectrum of the \lya\ transmission, we define the \lya\ flux contrast analogous to the density contrast, given by $\delta_\mathrm{F} = F / \bar{F} - 1$, where $\bar{F}$ is the mean flux.  The power pectrum is then computed as 
    \begin{equation}
        P_\mathrm{F,1D} (|k|) = v_\mathrm{scale} \langle \Tilde{\delta}_\mathrm{F} \Tilde{\delta}^*_\mathrm{F} \rangle,
    \end{equation}
    where $v_\mathrm{scale} = aH(z)L$, with $L$ being the length of the box, $\Tilde{\delta}_\mathrm{F}$ is the Fourier transform of $\delta_\mathrm{F}$, and the ensemble average is computed by averaging $\Tilde{\delta}_\mathrm{F} \Tilde{\delta}^*_\mathrm{F}$ over 5,000 sightlines for each $|k|$.

    We also perform a set of simulations with adaptive softening of gas particles.  This means that the gravitational softening length of gas particles is set by the SPH smoothing length.  As mentioned above, the default gravitational softening  length for gas particles in our fiducials simulations is the same as that of dark matter particles, and is set to $1/25$th of the mean interparticle separation \citep{2017MNRAS.464..897B}.  Adaptive softening can be enabled in \gadget\ by setting the \texttt{ADAPTIVE\_GRAVSOFT\_FORGAS} and \texttt{ADAPTIVE\_GRAVSOFT\_FORGAS\_HSML} flags.  The SPH smoothing kernel length is proportional to the comoving distance to the 32nd neighbour \citep{2001NewA....6...79S}.  As a result, at high redshifts, when the universe is more homogeneous, the softening length can be as large as three times the mean interparticle separation. As we will see below, this suppresses small-scale power in the gas density distribution. 
    
\section{Results}
\label{sec:results}

Our fiducial simulations consist of five runs with the five initial condition set-ups from Table~\ref{tab:code}.  The simulation specifications are as described above.  We now present the results from these simulations.  We will also occasionally refer to variations on these simulations.  There are four such variations that we consider.  One of these uses \ngenic\ with a glass-like initial particle load.  Another employs the same glass-based \ngenic\ set-up but with adaptive softening.  In all other respects, these runs are identical to our fiducial 20--256 runs.  Our third and fourth variations are the 20--512 and 20--1024 counterparts of our fiducial 20--256 runs.  These high-resolution runs have been mentioned above in Section~\ref{sec:setup}.

\begin{figure*}
  \begin{center}
    \includegraphics*[width=0.8\textwidth]{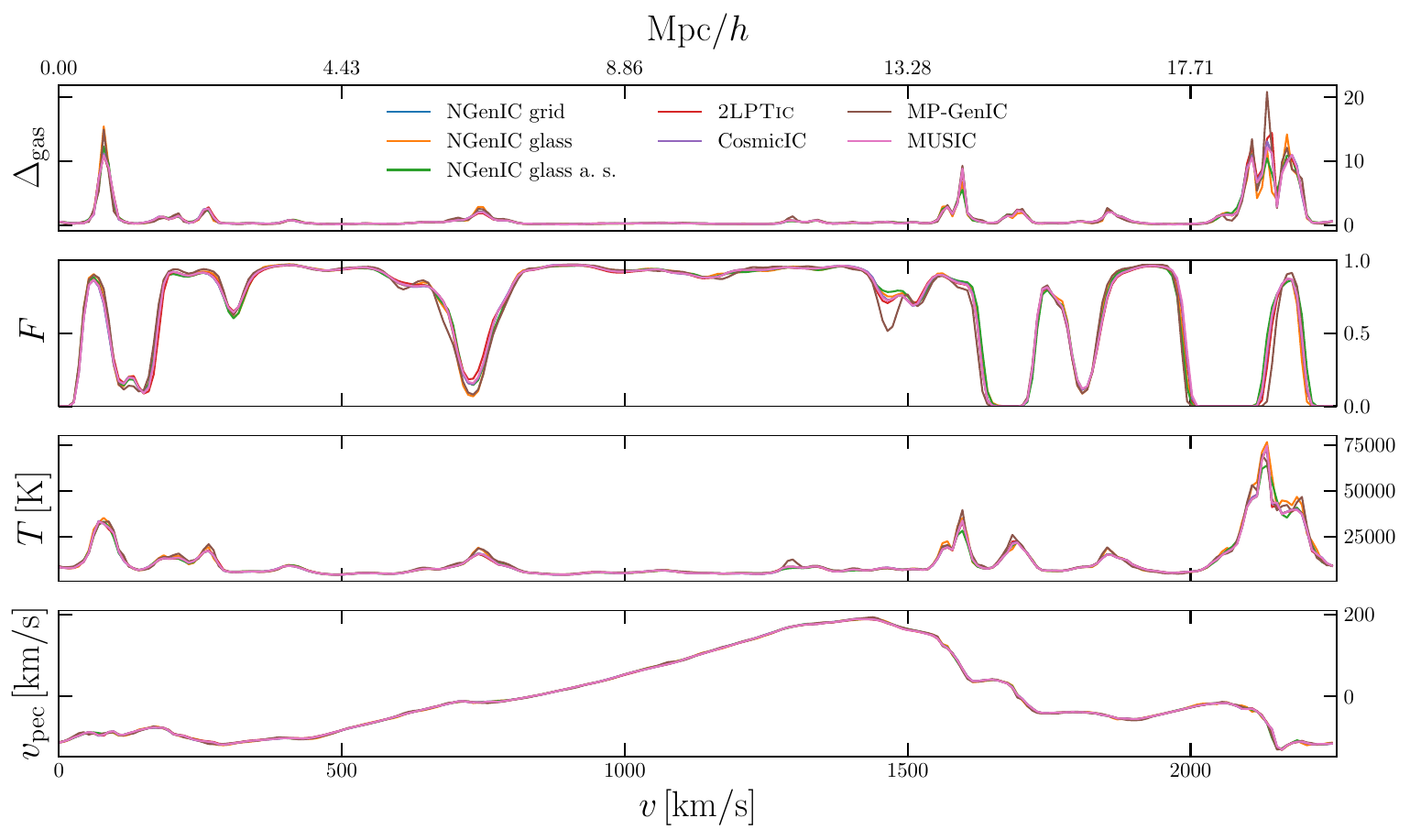}
  \end{center}
  \caption{The gas overdensity, Lyman-$\alpha$ transmitted flux, gas temperature and gas peculiar velocity along a randomly chosen one-dimensional skewer from our simulation boxes.  Each panel shows results from our five primary simulations.  Also shown are results from a glass-based \ngenic\ run, and a glass-based \ngenic\ run that uses adaptive softening (indicated by `a.~s.' in the legend).}
  \label{fig:skewers}
\end{figure*}

\begin{figure*}
 \begin{center}
    \includegraphics*[width=0.8\textwidth]{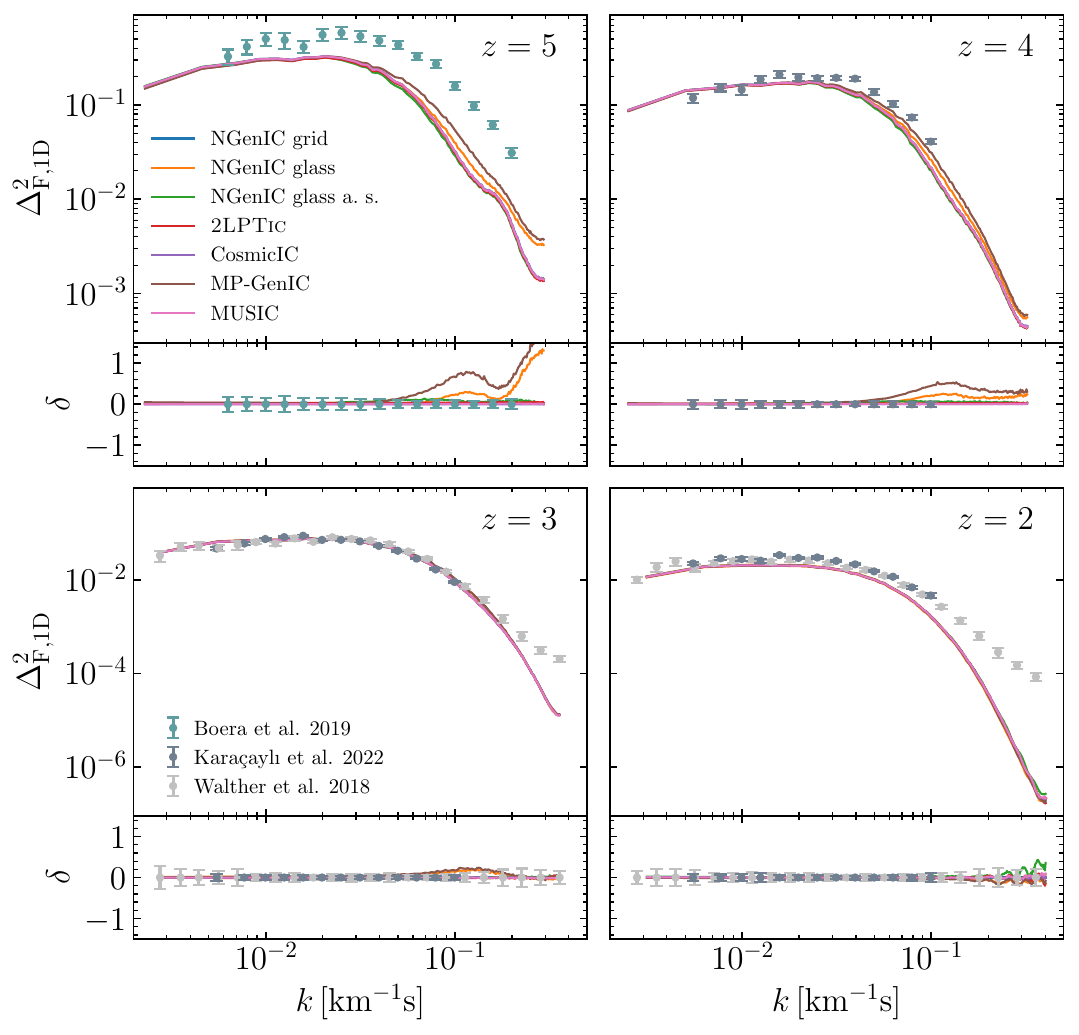}
  \end{center}
 \caption{The one-dimensional \lya\ transmission power spectrum at $z=2, 3, 4,$ and $5$.  Each panel shows results from our five primary simulations.  Also shown are results from a glass-based \ngenic\ run (orange curve), and a glass-based \ngenic\ run that uses adaptive softening (green curve, indicated by `a.~s.' in the legend).  We also show measurements by \citet{2018ApJ...852...22W}, \citet{2019ApJ...872..101B}, and \citet{2022MNRAS.509.2842K}.  These data are only shown for context and as a measure of the current uncertainties.  We make no attempt at fitting the data with our models, which explains the mismatch between the simulations and the observations at some redshifts and scales.  The error panels show differences relative to the result from the grid-based \ngenic\ simulation.  All spectra are normalised to the observed mean transmission from \citet{2013MNRAS.430.2067B}.}
 \label{fig:FluxPS}
\end{figure*}

For each of the six simulations shown in Figure~\ref{fig:InitialPS}, two-dimensional gas density distribution slices at $z=3$ are shown in Figure~\ref{fig:slices}.  Each slice has the thickness of a single grid cell, which in our case is 78.125~kpc$/h$.  The familiar non-linear large-scale structure at this redshift is clearly seen.  Furthermore, the structure in all simulations appears to be visually identical, showing that we are using identical random numbers while generating the initial conditions.  

\subsection{Nonlinear matter power spectrum}

The top four panels of Figure~\ref{dm_gas_ps} show the three-dimensional power spectrum of the dark matter density contrast at redshifts 2, 3, 4 and 5, in our five fiducial simulations.  We also show results from the simulation with glass-based \ngenic\ and the one with adaptive softening.  The linearly extrapolated matter power spectrum at respective redshifts is also shown.  At the bottom of each panel the difference relative to the grid-based \ngenic\ run is shown.  Compared to Figure~\ref{fig:InitialPS}, we can now see the power enhancement relative to linear theory at small scales due to nonlinear evolution.  As anticipated in Figure~\ref{fig:slices}, the power spectra agree between the different simulations to a large degree, but interesting differences are now visible.

Overall, all power spectra are consistent within 10\% over the scales shown in Figure~\ref{dm_gas_ps}.  The largest deviation from the fiducial run is seen in the \tlpt\ case, which exhibits more power relative to the \ngenic\ run.  The difference is at most $10\%$ at the scale of $k = 10\,h/$cMpc, and decays with time.  The dark matter distribution in our glass-based \ngenic\ run has less dark matter power at small scales relative to the \ngenic\ grid.  These differences appear to be enhanced when we use adaptive softening, but in either case the differences decay with time.  Our \cosmic\ results are nearly identical to those from the \ngenic\ run.  The \mpgenic\ and \music\ cases show only small differences that appear well below 5\% at all redshifts shown here. The difference between the fiducial run and the \music\ case can potentially be partly explained by the small difference between the value of the growth factor used in the two codes (see Section~\ref{sec:methods}).

The bottom four panels of Figure~\ref{dm_gas_ps} show the three-dimensional gas density power spectrum from the simulations shown in the top panel, along with the linear theory prediction, at redshifts 2, 3, 4 and 5.  As before, we show the relative difference from the grid-based \ngenic\ run at the bottom of each panel.  The simulations agree with each other reasonably well, with differences of at the most 10\% in the gas density power spectrum.  The greatest deviation is seen in the glass-based runs, \ngenic\ and \mpgenic. Both of these runs produce gas power larger by $\sim 10\%$ at small scales relative to the \ngenic\ using a grid.  The differences appear to increase towards the small scales resolved by the simulation.  This power enhancement is related to that seen in the initial conditions, shown in Figure~\ref{fig:InitialPS}, where we understood it to be the result of the nonlinearities in the process of creating the glass.  We also see that this power enhancement reduces with time as we go from $z=5$ to $2$ in Figure~\ref{dm_gas_ps}.  Further, the use of adaptive softening seems to remove the power excess, as noted previously by \citet{2012ApJ...760....4O}.  If we compare the dark matter and gas density power spectra in Figure~\ref{dm_gas_ps}, we can see that dark matter has more power than gas at all scales and all redshifts. This can be understood because of the exclusion of the gas particles that were turned into stars in biased high-density regions.  This results in the attenuation of power at small as well as large scales for the baryons \citep{2015ApJ...812...30K}.

\subsection{One-dimensional \lya\ transmission power spectrum}

We now turn to the \lya\ transmission properties of our simulations.  Figure~\ref{fig:skewers} shows one-dimensional sightlines, `skewers', from the simulations shown in Figure~\ref{dm_gas_ps}.  All sightlines are shown at $z=3$.  We show the gas density, the Lyman-$\alpha$ flux, gas temperature, and gas peculiar velocity along a randomly chosen sightline.  We normalise the simulations to identical mean transmission values.  We choose these to be the measurements by \citet{2013MNRAS.430.2067B}. As before, while the seven simulations appear very similar, there are small differences. These appear to be largest near the density peaks and troughs, with the \mpgenic\ run being the widest outlier.  The \lya\ transmission exaggerates these differences, so that there is up to a 30\% difference in the transmission.  Small differences also appear to be present in the gas temperature.  Most importantly, there are small differences in the peculiar velocity.  These small peculiar velocity differences will, however, have a large effect on the small-scale power spectrum of the \lya\ forest.  

Figure~\ref{fig:FluxPS} shows the one-dimensional \lya\ flux power spectrum from all the seven simulations at redshifts 2, 3, 4 and 5. We  show the dimensionless \lya\ transmission power spectrum.  At the bottom of each panel, we show differences relative to the grid-based \ngenic\ run.  The figure also shows measurements by \citet{2022MNRAS.509.2842K}, \citet{2018ApJ...852...22W}, and \citet{2019ApJ...872..101B}.  While we make no attempt the fit our simulations to these data, the uncertainty estimates on the measurements provide a context while examining the differences in our simulations.  A better agreement would require a well-calibrated thermal model run at higher resolution.  We now see that the glass-based runs using \ngenic\ and \mpgenic\ severely disagree with the grid-based result.  The differences are worse in the case of \mpgenic.  At $k=0.1$~s/km at $z=5$, \mpgenic\ can yield difference in excess of 50\%.  This is not only far in excess of the data uncertainty, but is also enough to significantly bias inferences of parameters such as the warm dark matter particle mass and the IGM temperature.  We will investigate the causes behind these differences below.  The differences decay with time to reduce to percent levels at $z=2$.  It is interesting to note that the result from the glass-based \ngenic\ model deviates less from the grid-based result than \mpgenic.  But both \mpgenic\ and the glass-based \ngenic\ runs show very similar $k$-dependence of the power spectrum, indicating a common origin for their behaviour. We also see that introducing adaptive softening in the glass-based \ngenic\ run removes the differences from its grid-based counterpart.  All other runs are in excellent agreement with each other.

\section{Understanding the small-scale differences}
\label{sec:discuss}

\begin{figure}
  \begin{center}
    \includegraphics*[width=\columnwidth]{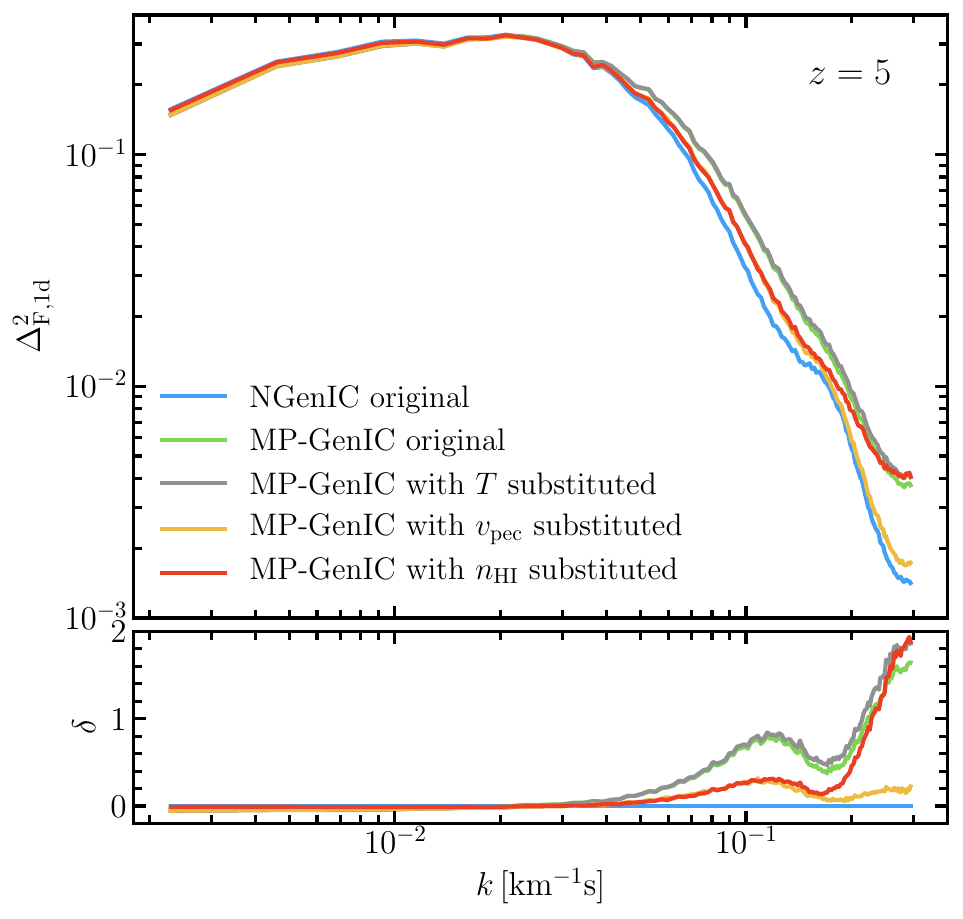}
    \caption{The one-dimensional \lya\ transmission power spectrum from our \ngenic\ and \mpgenic\ runs at $z=5$.  The blue and green curves show our fiducial grid-based \ngenic\ and \mpgenic\ runs, respectively.  The other curves show the power spectrum from the \mpgenic\ run when the temperatures (grey curve), peculiar velocities (yellow curve), or gas densities (red curve) in this simulation are substituted by their corresponding values from the \ngenic\ run.  The simulations are normalised to the observed mean flux in all cases, as described in the text.  The bottom panel shows the fractional change with respect to the \ngenic\ result.}
    \label{fig:mpgenic_swap_test}
  \end{center}
\end{figure}

\begin{figure}
  \begin{center}
    \includegraphics*[width=\columnwidth]{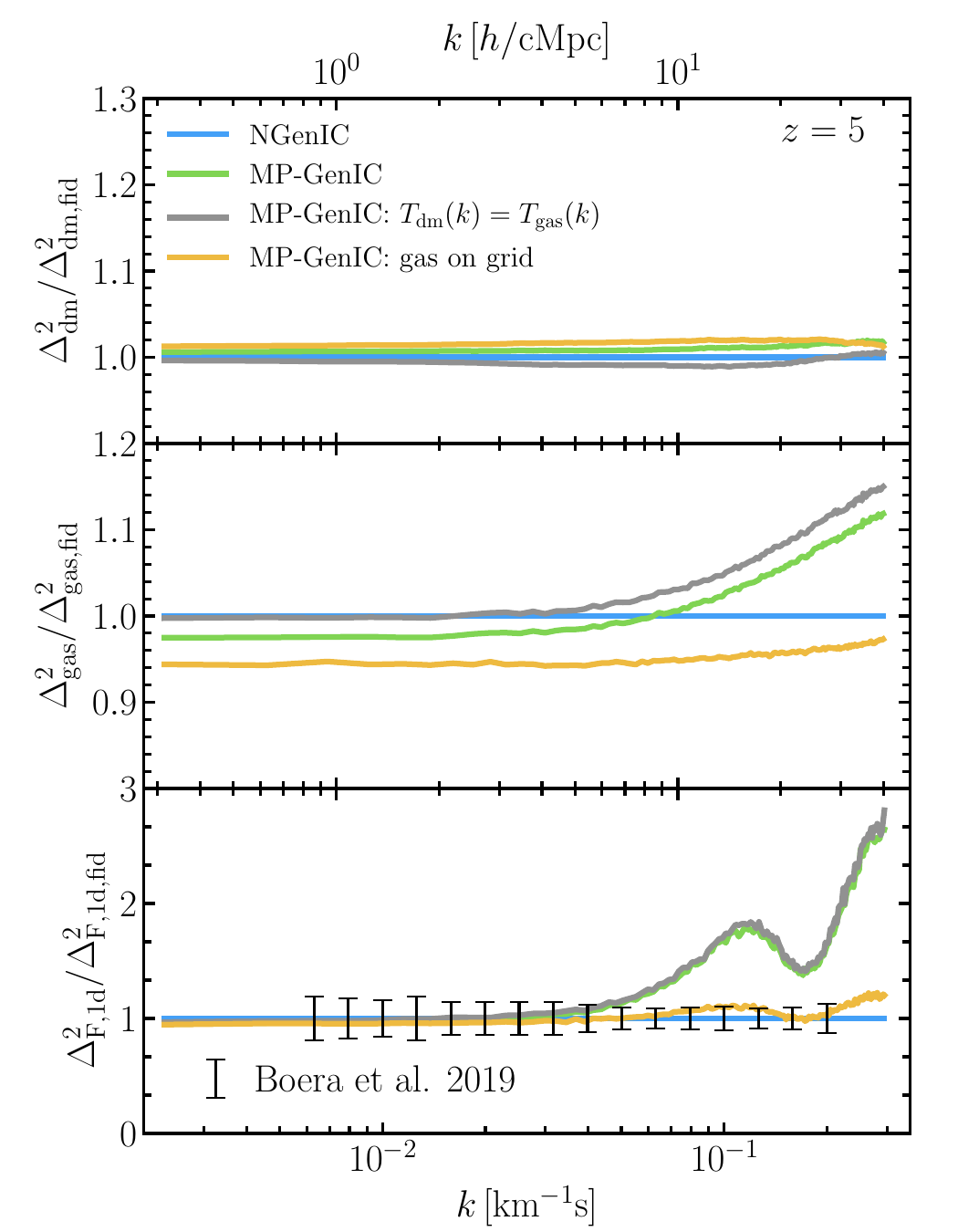}
    \caption{Effect of modifying the gas particle set-up in our \mpgenic\ run on the dark matter density power spectrum (top panel), gas density power spectrum (middle panel), and the one-dimensional \lya\ flux power spectrum.  The blue and green curves show our fiducial grid-based \ngenic\ and \mpgenic\ runs, respectively.  The grey curves show results from an \mpgenic\ run in which identical transfer functions are used for gas and dark matter density perturbations.  The yellow curves show results from an \mpgenic\ run with different transfer functions but with the unperturbed initial positions of gas particles as well as dark matter particles following a uniform grid.  The error bars in the bottom panel are uncertainties on the power spectrum measurements by \citet{2019ApJ...872..101B}.}
    \label{fig:mpgenic_test}
  \end{center}
\end{figure}

\begin{figure}
  \begin{center}
    \includegraphics*[width=\columnwidth]{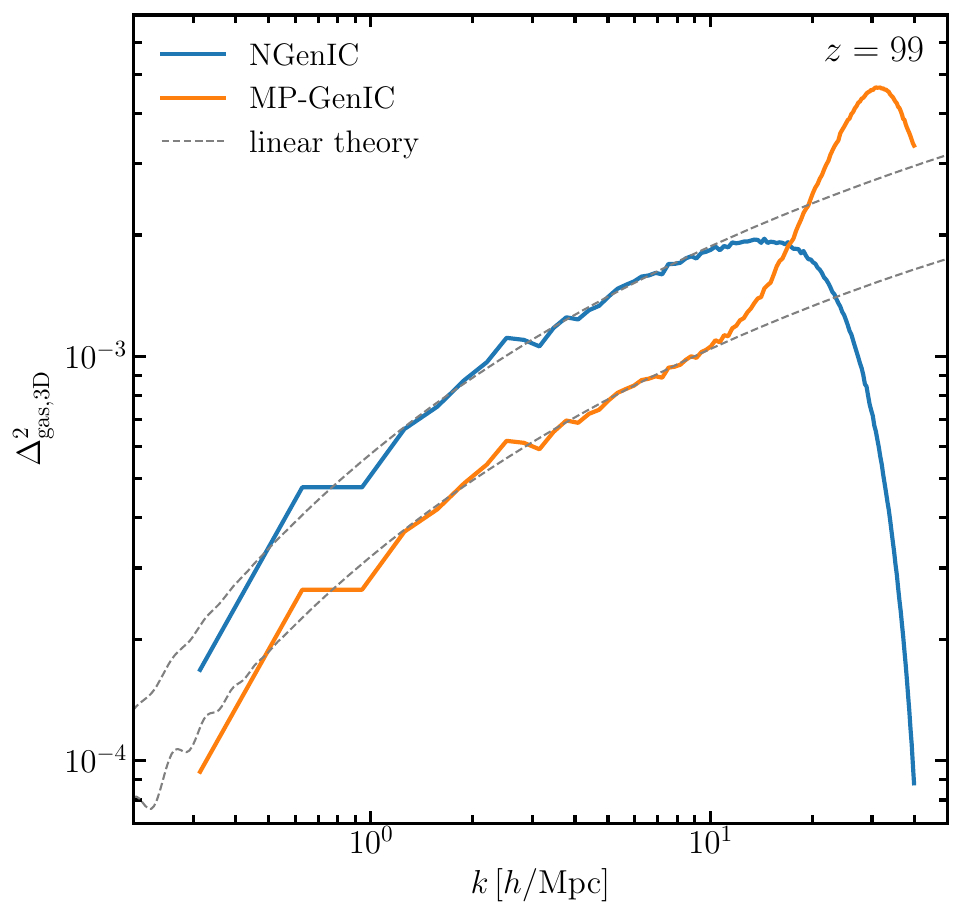}
  \end{center}
  \caption{The three-dimensional power spectrum of the gas density at $z=99$ in our \ngenic\ and \mpgenic\ simulations.  The difference in the overall amplitude is due to the difference in the transfer functions used in the two runs.  The differences at the small scales are due to the use of glass in \mpgenic.  The dashed curves show the linearly extrapolated power spectra in the two cases.}
  \label{diff_ic_gas_power}
\end{figure}

\begin{figure}
  \begin{center}
    \includegraphics*[width=\columnwidth]{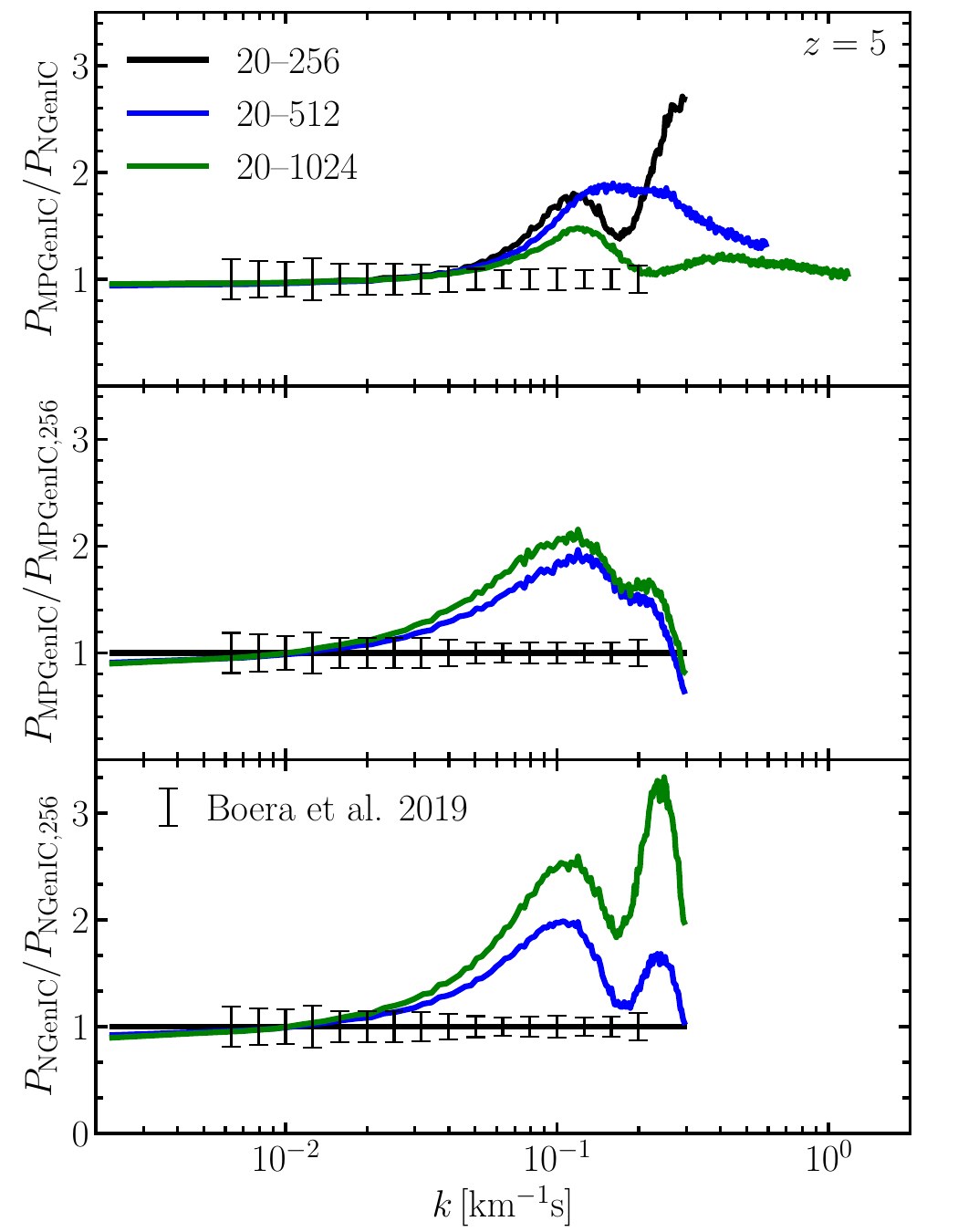}
  \end{center}
  \caption{The top panel shows how the \lya\ flux power spectrum ratio between \ngenic\ and \mpgenic\ changes  with resolution. The middle panel shows the change in the \mpgenic\ run with resolution. The bottom panel shows the same for our \ngenic\ run. Even at the highest resolution, there is a $\sim 50\%$ difference in the \lya\ flux power spectrum from the two runs.  The error bars in each panel are uncertainties on the power spectrum measurements by \citet{2019ApJ...872..101B}. These are shown simply for a comparison with the differences between the simulation curves.}
  \label{fig:20_1024}
\end{figure}

Figure~\ref{fig:mpgenic_swap_test} helps us understand the reasons behind the large difference that we see between the \lya\ flux power spectrum from our \ngenic\ and \mpgenic\ runs.  This figure shows the \lya\ flux power spectrum at $z=5$.  The green and blue curves show our fiducial results from \mpgenic\ and \ngenic, respectively.  These power spectra are the same as those in Figure~\ref{fig:FluxPS} above.  The other curves show the power spectrum from the \mpgenic\ run when the temperatures (grey curve), peculiar velocities (yellow curve), or gas densities (red curve) in this simulation are substituted by their corresponding values from the \ngenic\ run.  The simulations are normalised to the observed value of the mean flux in all cases.  The bottom panel shows the fractional change with respect to the \ngenic\ result.  We see that there is no significant difference between the gas temperatures in the two simulations.  When the gas temperature in the \mpgenic\ run is substituted by the \ngenic\ run, the \lya\ flux PS shows almost no change.  But the effect is significant when we substitute the gas velocity or the gas density.  Gas velocity is the greater effect between the two.  At $k> 0.2$~s/km, almost the entire difference between the power spectra comes from gas velocity.

As Figure~\ref{fig:mpgenic_test} shows this behaviour is entirely due to the numerical choice of using a glass-like configuration for the gas particles in our \mpgenic\ run.  The figure shows the relative change in the dark matter power spectrum (top panel), the gas density power spectrum (middle panel), and the one-dimensional \lya\ flux power spectrum.  Recall that there are two important differences between our \ngenic\ and \mpgenic\ simulations: (\textit{a}\/) the gas and dark matter initial power spectra are identical in the \ngenic\ run, but these are different in the \mpgenic\ run, and (\textit{b}\/) gas particles are initialised on a grid in the \ngenic\ run whereas they are initialised on a glass in the \mpgenic\ run.  Figure~\ref{fig:mpgenic_test} investigates what happens when we remove each of these differences between the two runs one at a time.  The blue and green curves in this figure show various power spectra from our fiducial \ngenic\ and \mpgenic\ runs.  The grey curve shows power spectra from an \mpgenic\ run in which we use identical initial transfer functions for gas and dark matter.  The yellow curves show power spectra for an \mpgenic\ run in which the gas and dark matter transfer functions are distinct as before but now the gas particle positions are initialised on a grid.  We see that the dark matter power spectra are nearly identical in all cases.  The small differences of $\lesssim 2$\% result partly from the choice between the use of the total matter power spectrum and dark matter power spectrum, but also due to a backreaction of the large differences in the gas power spectra. However, the gas density power spectra are significantly different.  Here, we first see that there is a broadband difference between the \ngenic\ and \mpgenic\ runs because of the difference in the transfer functions.  This difference vanishes at large scales when we use the same transfer functions for gas and dark matter in the \mpgenic\ run.  Next, at small scales, the order 10\% differences in the gas density power spectra are due to the adoption of glass-like initial conditions.  We see that this difference goes away as we shift from glass to grid.  Figure~\ref{diff_ic_gas_power} shows that these differences in the gas power spectra are already present at $z=99$.  (In Figure~\ref{diff_ic_gas_power}, we also show the linearly extrapolated power spectra for the two simulations.  We see that both simulations show deviation from the linear theory prediction at small scales.  The \ngenic\ run underpredicts the small-scale power.  This is the result of smoothing due to the SPH kernel.  The \mpgenic\ run overpredicts the small-scale power.  This is because of the use of glass.) Finally, turning to the \lya\ flux, the bottom panel of Figure~\ref{fig:mpgenic_test} shows that the \lya\ flux power spectrum does not change at all with the change in the gas initial power spectrum.  However, it shows a big change when we shift from glass to grid.  Indeed, when we used a grid in our \mpgenic\ run, the flux PS is nearly identical to the \ngenic\ result.  We are thus in a situation that the set-up designed to reproduce the correct linear-theory behaviour ends up introducing significant changes in the \lya\ transmission power spectra.

All of the above results used our default 20--256 runs.  However, this spatial resolution has been known to be insufficient for convergence for the \lya\ flux PS at $z=5$ \citep{2009MNRAS.398L..26B, 2021PhRvL.126g1302R, 2023MNRAS.tmp.2467D}.  In Figure~\ref{fig:20_1024}, we look into the effect on the flux power spectrum of the spatial resolution of our runs.  We now do our \ngenic\ and \mpgenic\ runs at 20--512 and 20--1024 resolutions.  In each case, we set the value of \texttt{PMGRID} to the cube root of the particle number, and we use a softening length that is 1/25 of the mean inter-particle spacing.  See Appendix~\ref{sec:soft} for how these results change if we change these settings.  Note that 20--1024 is the resolution used by many \lya\ forest models in the literature.  We see in Figure~\ref{fig:20_1024} that even at our highest resolution there is upto 50\% difference in the flux PS at $z=0.1$~s/km between our \ngenic\ run and \mpgenic\ runs.  It is also important to note that the difference is scale dependent, thus precluding the possibility of applying simple scale-independent `resolution corrections' to low-resolution runs.  (The enhancement in power at $k\sim 0.25\;$s/km, is similar in shape to the enhancement that \citet{2023arXiv230904533I} find when they increase the cumulative energy input per proton mass at mean energy, $u_0$.  This suggests that heat injection might also be playing a role in the behaviour of the power spectrum in Figure~\ref{fig:20_1024}.  The value of $u_0$ in our simulations is likely to be higher than that in the reference simulation of \citet{2023arXiv230904533I} due to earlier reionization in our models.)

\section{Conclusions}
\label{sec:conclusions}

We have examined the effects of various assumptions and algorithms used in setting up the cosmological initial conditions for cosmological hydrodynamic simulations that are used to model the \lya\ forest.  Our goal was to understand if the predictions made by these models for the \lya\ forest are robust against algorithmic choices and other assumptions.  For this purpose, we designed simulations that were identical in all other respects, including in the manner of sampling the Gaussian initial density distribution.  We considered five initial conditions codes, and studied four further variations of some of these codes.  Our conclusions are as follows:

\begin{itemize}
\item At the high redshifts typically used to initialise cosmological simulations, when initial conditions codes employ a grid configuration for the initial unperturbed particle positions, their resultant perturbed density fields have a three-dimensional matter density power spectrum that agrees with the linear theory down to the smallest scales resolved in our simulations.  This is not the case, however, when a glass-like configuration is used for the initial unperturbed particle positions.  In these glass-based cases, the matter density power spectrum shows a significant enhancement at small scales relative to linear theory.  This is also the case when a glass-like configuration is used only for gas particles.

\item The non-linear dark matter and gas density power spectra, at redshifts $z=5$--$2$, agree to within 10\% in all our runs.  For the dark matter density, the use of second-order Lagrangian perturbation theory causes the largest differences, relative to runs that use the Zel'dovich approximation.  The use of adaptive softening can also lead to large differences in the dark matter density power spectrum.  For gas, the largest difference is seen when a glass-like initial particle load is used. There are also large broadband differences when separate transfer functions are used for gas and dark matter.  All of these differences, in dark matter and gas alike, decay with time, and are relatively insignificant at $z<3$.  The small-scale power enhancement can be offset by using adaptive softening.

\item We report significant differences in the one-dimensional \lya\ transmission power spectrum between our grid-based and glass-based runs.  At $k\sim 0.1$~s/km, the flux power spectrum has differences of the order of 50\% between our grid-based \ngenic\ run and our \mpgenic\ run that uses a glass-like initial position for gas particles.  These differences are scale-dependent, and therefore have a significant potential to bias any cosmological inferences made using similar models.  These differences also appear to hold at high resolution.  The source of the difference appears to be the choice of glass-like particle loads, which affects the small-scale gas velocities and densities. Of these two runs, only the \mpgenic\ run correctly reproduces the expected linear theory behaviour in the difference of power between gas and dark matter densities.
\end{itemize}

It is not clear to us how important it is for models of the small-scale structure of the \lya\ forest to reproduce the linear theory behaviour of the power difference between gas and dark matter densities.  Nonetheless, the means of achieving this via the use of glass-like particle configurations appear to lead to large distortions in the \lya\ flux power spectrum.
These distortions can be considered a numerical artefact, for two reasons.  First, these distortions appear similar to the distortions seen in fully glass-based runs, in which both gas and dark matter particles are initialised on a glass, and where this behaviour is understood to be a numerical feature.  Second, the matter density power spectrum in these runs fails to agree with the linear theory at even high redshifts.  For all other things kept the same, this distortion causes an enhancement in the flux power spectrum at small scale.  This can bias inferences in parameters such as the axion mass or the warm dark matter particle mass, and can potentially explain some of the very strict limits recently reported in the literature. In this context, it is interesting to note that \citet{2021MNRAS.503..426H} have also proposed a method to achieve the correct linear theory growth of the difference in the gas and dark matter density power spectra. In their approach the correct linear theory evolution is obtained by implementing the transfer functions as mass perturbation to each species, thus allowing the use of grids for both the dark matter and gas. This presents a potentially interesting alternative to consider in future efforts in the developing more robust models of the Lyman-$\alpha$ forest.

\section*{Acknowledgments}

We thank Simeon Bird, Frederick Davies, Prakash Gaikwad, Joseph Hennawi, Zarja Lukic, and Rishi Khatri for useful discussions. We also thank the anonymous referee for their very helpful comments. GK gratefully acknowledges support by the Max Planck Society via a partner group grant. GK is also partly supported by the Department of Atomic Energy (Government of India) research project with Project Identification Number RTI 4002.  This work used the Cambridge Service for Data Driven Discovery (CSD3) operated by the University of Cambridge (www.csd3.cam.ac.uk), provided by Dell EMC and Intel using Tier-2 funding from the Engineering and Physical Sciences Research Council (capital grant EP/P020259/1), and DiRAC funding from the Science and Technology Facilities Council (www.dirac.ac.uk).  This work further used the COSMA Data Centric system operated by Durham University on behalf of the STFC DiRAC HPC Facility. This equipment was funded by a BIS National E-infrastructure capital grant ST/K00042X/1, DiRAC Operations grant ST/K003267/1 and Durham University. DiRAC is part of the UK's National E-Infrastructure.  JSB is supported by STFC consolidated grant ST/T000171/1. Support by ERC Advanced Grant 320596 `The Emergence of Structure During the Epoch of Reionization’ is gratefully acknowledged. MGH has been supported by STFC consolidated grant ST/N000927/1 and ST/S000623/1. VI is supported by the Kavli Foundation. 

\section*{Data availability}
No new data were generated or analysed in support of this research.

\bibliographystyle{mnras}
\bibliography{Biblo}

\appendix

\section{Particle offsets in initial conditions}

\begin{figure}
  \begin{center}
    \includegraphics*[width=\columnwidth]{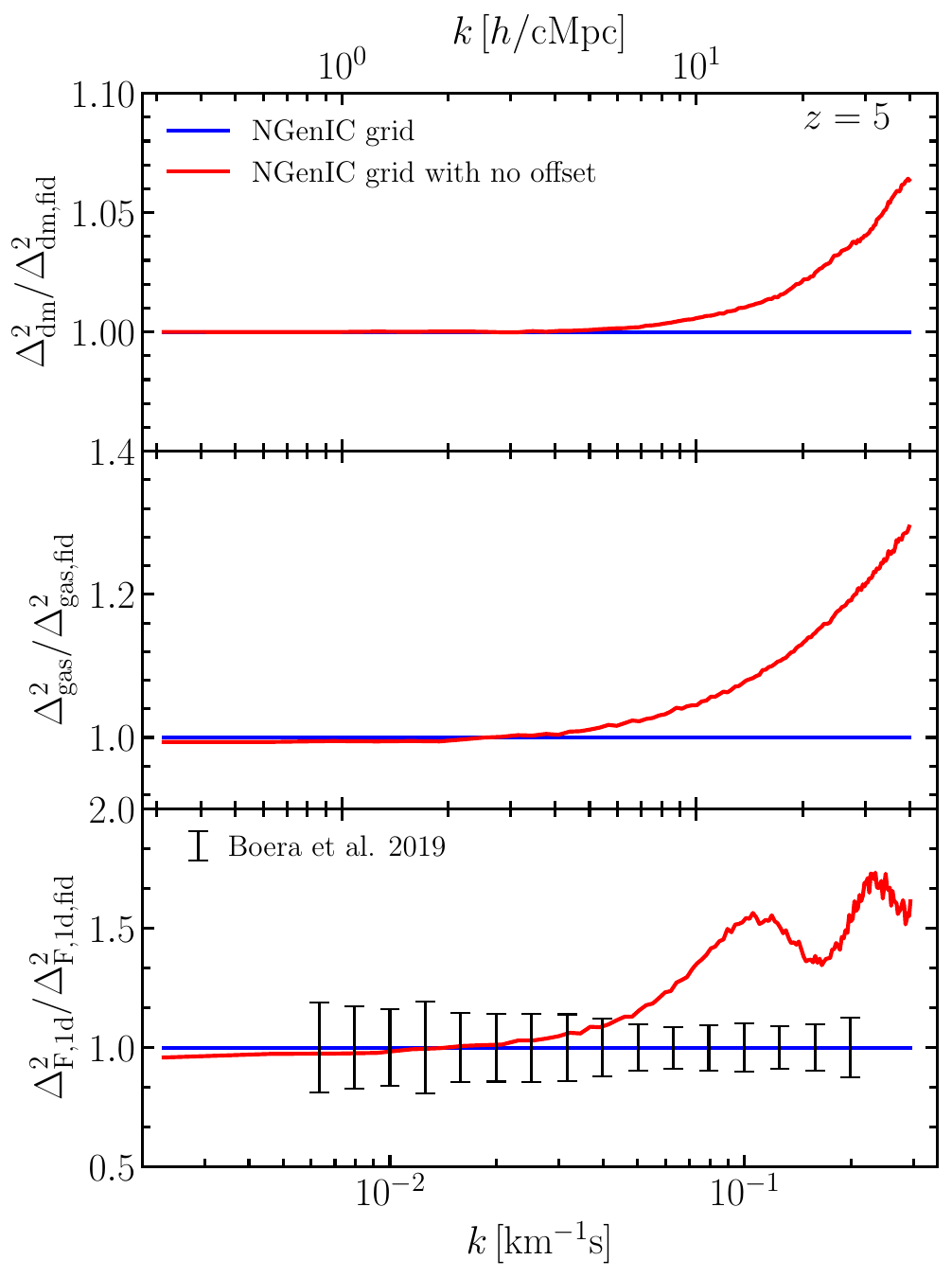}
  \end{center}
  \caption{Effect of using a position offset in the initial gas and dark matter particle positions on the three-dimensional dark matter density power spectrum (top panel), the three-dimensional gas density power spectrum (middle panel), and the one-dimensional \lya\ transmission power spectrum (bottom panel).  We plot the ratio of the power spectrum from a simulation that does not offset the particle positions, thus placing gas particles at the same positions as the dark matter particles while starting the simulation.  In the blue curve, the default offset, explained in Section~\ref{sec:ngenic}, is used.  Both simulations are 20--256 and use \ngenic.}
  \label{fig:noOffset}
\end{figure}

As discussed above, in Section~\ref{sec:ngenic}, we introduce an offset in gas particle positions relative to dark matter particle positions in our \ngenic\ run.  This is done to avoid spurious small-scale power due to gravitational coupling between the two species of particles.  Figure~\ref{fig:noOffset} shows the effect that not offsetting the particles has on the dark matter density power spectrum, the gas density power spectrum, and the one-dimensional \lya\ transmission power spectrum.  Only redshift $z=5$ is shown for illustration.  As noted by \citet{2012ApJ...760....4O}, gravitational coupling between gas and dark matter particles can lead to a significant excess power at small scales. It is interesting to note that particle offsets in a grid-based \ngenic\ run produce changes in the transmission power spectrum that are similar to the changes introduced when one goes from the grid-based \ngenic\ run with offset to the \mpgenic\ run (Figure~\ref{fig:FluxPS}). This suggests that the differences between \mpgenic\ and \ngenic\ could be caused by particle coupling and two-body interactions.

\section{Effect of radiation}
\label{app:rad}

\begin{figure}
  \begin{center}
    \includegraphics*[width=\columnwidth]{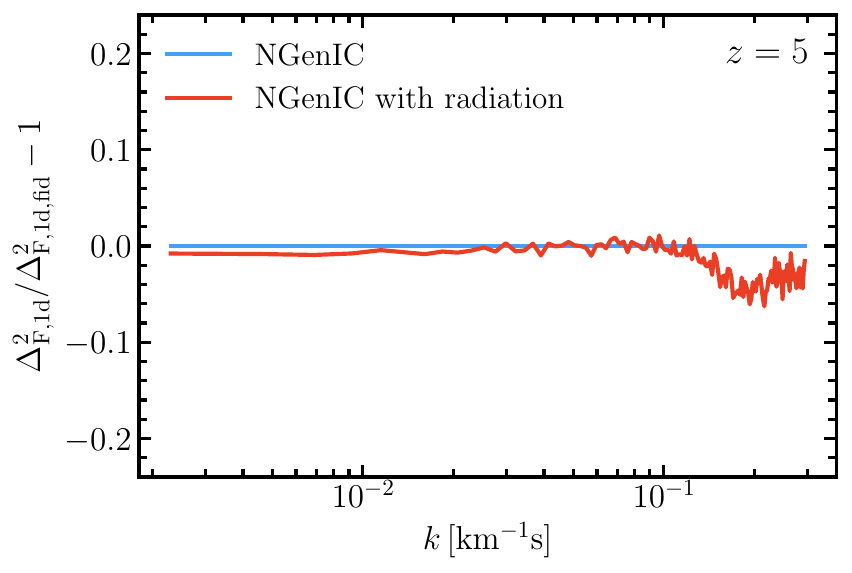}
  \end{center}
  \caption{The one-dimensional \lya\ transmission power spectrum at $z=5$ in a simulation that does not account for the radiation density term (blue curve) and one that does account for it (red curve).  Both simulations use \ngenic\ with an initial power spectrum derived using CLASS and normalised to give $\sigma_8=0.829$ at $z=0$.}
  \label{fig:PSComparison_z5_radiation}
\end{figure}

We use CLASS to obtain the initial power spectrum at $z=99$.  This input power spectrum is normalised to obtain our desired value of $\sigma_8=0.829$ at $z=0$.  This normalisation is done using a growth factor that does not include the radiation term, in order to be consistent with our use of the \gadget\ code, which also does not include the radiation term while evolving the density perturbations. Figure~\ref{fig:PSComparison_z5_radiation} shows the effect that this has on the \lya\ flux power spectrum relative to a \gadget\ run that uses the radiation term for the background cosmological evolution for an identical input power spectrum.  The figure shows the power spectrum from two 20--256 simulations using \gadget, one of which does not use the radiation density term (blue curve) and one that does use it (red curve).  Both simulations use our default grid-based \ngenic\ set-up to compute the initial conditions, so that the input power spectrum is normalised in identical fashion in both cases.  We see that an inconsistency in incorporating the radiation term leads to a small bias of less than 10\% in the power at $k\sim 0.1$~s/km.  There is no significant effect at larger scales.  

\section{Starting redshift}
\label{sec:zstart}

\begin{figure}
  \begin{center}
    \includegraphics*[width=\columnwidth]{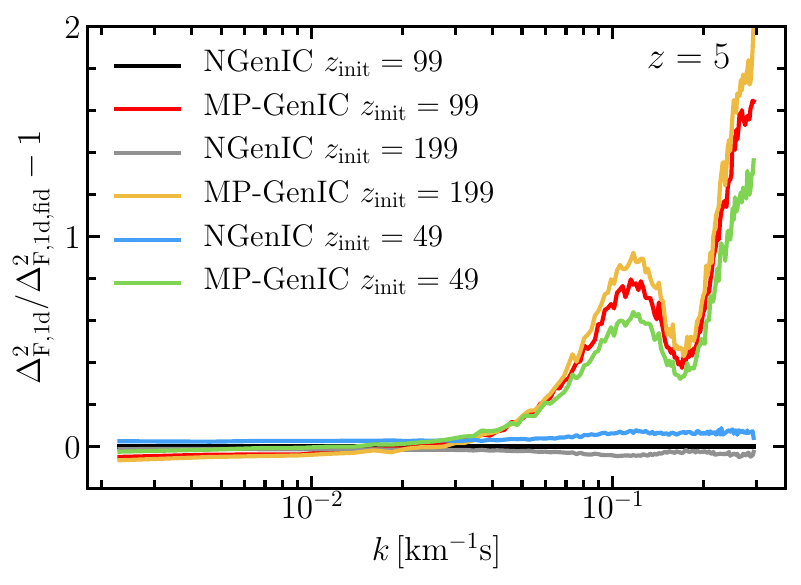}
  \end{center}
  \caption{Dependence of our results on the starting redshift of the simulations.  The black and red curves show results from our default \ngenic\ and \mpgenic\ simulations, respectively, both of which start at $z_\mathrm{init}=99$.  The grey and orange curves show results from the simulation when they are initialised at $z_\mathrm{init}=199$ instead.  The blue and green curves show the same for $z_\mathrm{init}=49$. Changing the initialisation redshift does not have a large effect on the flux power spectrum.} 
  \label{fig:zinit}
\end{figure}

Figure~\ref{fig:zinit} shows the effects of changing the starting redshift of our simulations from $z=99$ to $z=199$ and $z=49$.  This has a very small impact on the result. The small differences that do exist seem to indicate that the difference between the  power spectra in the \mpgenic\ and \ngenic\ cases are smaller for a smaller starting redshift. This observation, combined with the similarity in the shape of the distortion seen in the bottom panel of Figure~\ref{fig:noOffset} with that in Figure~\ref{fig:mpgenic_swap_test} and other related figures, suggests that the distortions may arise at least partly due to particle coupling. When the simulation is started at a later redshift there is relatively less time for the accumulation of particle coupling error, which explains the behaviour seen in Figure~\ref{fig:zinit}.

\section{Softening length convergence test}
\label{sec:soft}

\begin{figure*}
  \begin{center}
    \includegraphics*[width=0.7\textwidth]{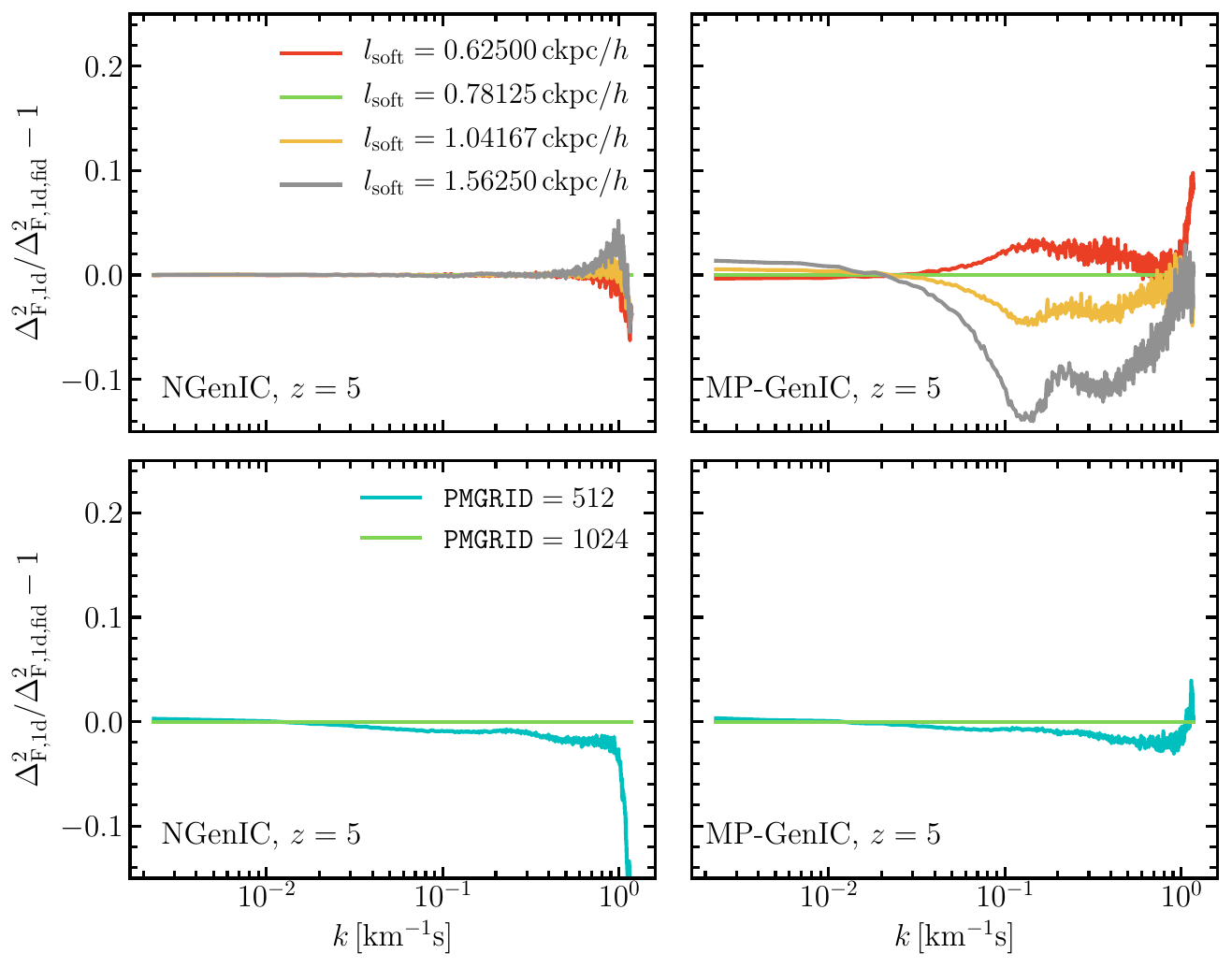}
  \end{center}
  \caption{Gravitational softening length and \texttt{PMGRID} convergence tests for \ngenic\ and \mpgenic\ at redshift $z=5$. All four panels show the change in the one-dimensional \lya\ flux power spectrum with respect to one of the default run. This is shown by the green curves in all the panels, and represents our default choice of $l_{\mathrm{soft}} = 0.78125$\,ckpc/$h$ and $\texttt{PMGRID} = 1024$. The left panels show the results for the \ngenic\ and the right panels are for the \mpgenic\ initial condition generators. The red, orange and grey curves show the results for softening lengths of 0.80, 1.33 and 2 times the default value keeping the \texttt{PMGRID} value to the default. From the upper two panels, we can see that for \ngenic\ the power spectrum is converged with respect to the softening length, but that from the \mpgenic\ had not. \mpgenic\ produces more power with reduced softening. In the lower panels, we use $\texttt{PMGRID} = 512$ for the cyan-colored curves. The \lya\ flux power spectrum does not change significantly with \texttt{PMGRID} in either of the codes.}   
  \label{fig:pmgrid_lsoft}

\end{figure*}

\begin{figure*}
  \begin{center}
    \includegraphics*[width=\columnwidth]{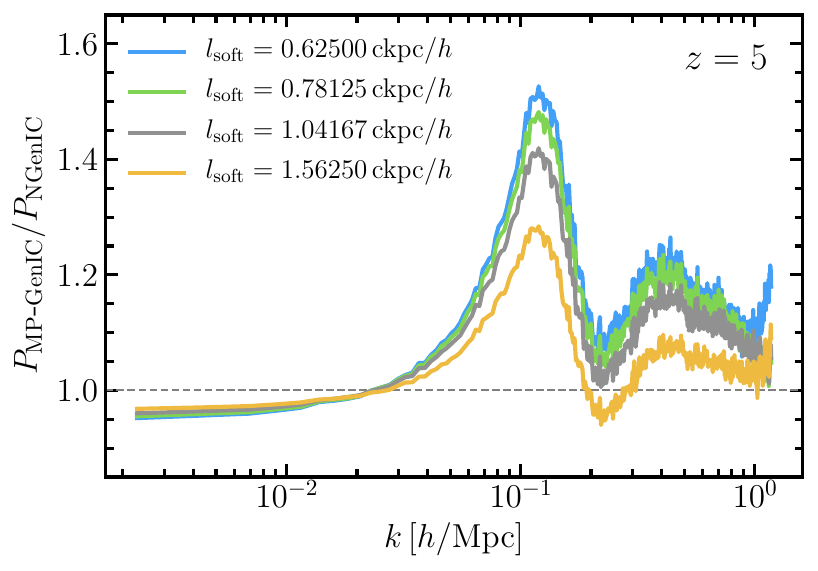}
  \end{center}
  \caption{Dependence of the \lya\ flux power ratio between \mpgenic\ and \ngenic\ on the gravitational softening length. We keep \texttt{PMGRID} fixed to the fiducial value of 1024 in all the curves here. The default softening length is shown by the green curve. The blue, grey and orange curves are for softening lengths of 0.8, 1.33 and 2 times the default length. Reducing the softening length increases the differences between the two codes.}  
  \label{fig:pmgrid_lsoft_test}
\end{figure*}

Figures~\ref{fig:pmgrid_lsoft} and \ref{fig:pmgrid_lsoft_test} show how the result in Figure~\ref{fig:20_1024} depends on the choice of  the values of the \texttt{PMGRID} and $l_\mathrm{soft}$ parameters.  The \texttt{PMGRID} parameter controls the resolution of the mesh used in the PM force calculation.  The softening length $l_\mathrm{soft}$ controls the short-range interaction between particles.  The bottom panels of Figures~\ref{fig:pmgrid_lsoft} show that the results of our 20--1024 simulation are not very sensitive to the value of \texttt{PMGRID}.   The same is true for the value of $l_\mathrm{soft}$ for the \ngenic\ run.  However, the \mpgenic\ run shows significant changes of the order of 10\% in the flux power spectrum.  Figure~\ref{fig:pmgrid_lsoft_test} shows that the difference in the power spectra in the two simulations worsens with decreasing softening length. The reduction in the difference between \ngenic\ and \mpgenic\ with increasing softening length suggests that the differences could be caused by particle coupling and two-body interactions.

\section{One-dimensional \lya\ transmission PDF}

\begin{figure*}
 \begin{center}
   \includegraphics*[width=0.75\textwidth]{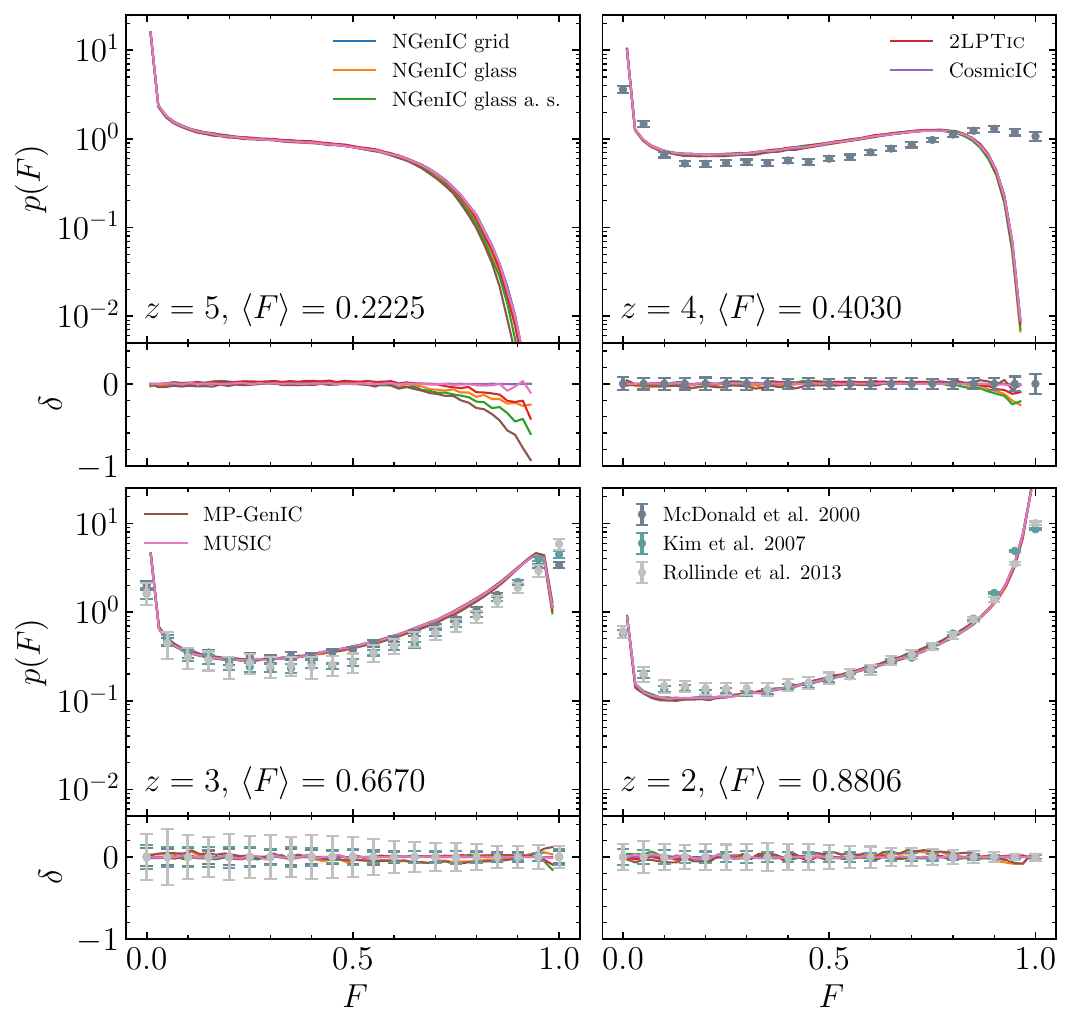}
  \end{center}
 \caption{The PDF of \lya\ transmission at $z=2, 3, 4,$ and $5$.  Each panel shows results from our five primary simulations.  Also shown are results from a glass-based \ngenic\ run, and a glass-based \ngenic\ run that uses adaptive softening (indicated by `a.~s.' in the legend).  We also show measurements by \citet{2000ApJ...543....1M}, \citet{2007MNRAS.382.1657K}, and \citet{2013MNRAS.428..540R}.  These data are only shown for context and as a measure of the current uncertainties.  We make no attempt at fitting the data with our models, which explains the mismatch between the simulations and the observations at some redshifts and transmission values.  The error panels show differences relative to the result from the grid-based \ngenic\ simulation.  All spectra are normalised to the observed mean transmission, the values of which are given in the respective panels.} 
  \label{fig:FluxPDF}
\end{figure*}

The \lya\ flux probability distribution function (PDF) is shown in Figure~\ref{fig:FluxPDF}, with the differences relative to the grid-based \ngenic\ run plotted at the bottom of each panel. We also show measurements of the PDF by \citet{2000ApJ...543....1M}, \citet{2007MNRAS.382.1657K}, and \citet{2013MNRAS.428..540R}.  To compute the PDF, we took 50 equal-sized bins between 0 and 1.  At $z = 4$ and $5$, the flux PDF peaks at $F=0$ but at lower redshifts the mode moves to $F=1$ as the forest becomes increasingly transparent.  Like all the other statistics, the difference between the simulations for \lya\ transmission PDF decreases with time.  The differences are highest at $z = 5$.  At these redshifts the PDF from the \mpgenic\ run shows almost 100\% difference near $F=1$ from that in the grid-based \ngenic\ run.  The glass-based \ngenic\ run is discrepant too, but with smaller differences, of about 20\% at $F = 1$.  Interestingly, using adaptive softening seems to enhance these differences.  There is good agreement between the rest of the simulation, except perhaps in the case of \tlpt, which also shows about 20\% differences, most likely as a result of the higher accuracy of the density perturbation computation.  

\bsp	
\label{lastpage}
\end{document}